\documentclass{article}

\usepackage[english]{babel}

\usepackage[letterpaper,top=2cm,bottom=2cm,left=3cm,right=3cm,marginparwidth=1.75cm]{geometry}
\usepackage[font=small,labelfont=bf,margin=0.5cm]{caption}

\usepackage[square,numbers]{natbib}

\usepackage{amsmath}
\usepackage{graphicx}
\usepackage[colorlinks=true, allcolors=blue]{hyperref}
\usepackage{subcaption}

\usepackage{amsfonts}
\usepackage{booktabs}
\usepackage{siunitx}

\title{\bf\Large Phases of Tree-decorated Dynamical Triangulations in 3D}
\author{\textsc{Timothy Budd} \, and \, \textsc{Dániel Németh} \\[5mm]
{\small IMAPP, Radboud University, Nijmegen, The Netherlands.}\\
{\small \texttt{\href{mailto:t.budd@science.ru.nl}{t.budd@science.ru.nl}, \href{mailto:nemeth.daniel.1992@gmail.com}{nemeth.daniel.1992@gmail.com}}}}

\begin{document}
\maketitle

\begin{abstract}
This work revisits the Euclidean Dynamical Triangulation (DT) approach to non-perturbative quantum gravity in three dimensions. Inspired by a recent combinatorial study \cite{budd2022family} of a subclass of 3-sphere triangulations constructed from trees, called the \emph{triple-tree} class, we present a Monte Carlo investigation of DT decorated with a pair of spanning trees, one spanning the vertices and the other the tetrahedra of the triangulation. The complement of the pair of trees in the triangulation can be viewed as a bipartite graph, called the \emph{middle graph} of the triangulation. In the triple-tree class, the middle graph is restricted to be a tree, and numerical simulations have displayed a qualitatively different phase structure compared to standard DT.
Relaxing this restriction, the middle graph comes with two natural invariants, namely the number of connected components and loops. Introducing corresponding coupling constants in the action, allows one to interpolate between the triple-tree class and unrestricted tree-decorated DT. Simulations of this extended model confirm the existence of a new phase, referred to as the \emph{triple-tree phase}, besides the familiar crumpled and branched polymer phases of DT. A statistical analysis of the phase transitions is presented, showing hints that the branched polymer to triple-tree phase transition is continuous.

\end{abstract}

\section{Introduction}

Building upon the ideas introduced by Weinberg \cite{Weinberg:1980gg} and hypothesizing the existence of an ultraviolet fixed point for gravity, a sensible non-perturbative approach to Euclidean quantum gravity is to seek self-similar scaling limits in discretized gravity models. 
One of these approaches, \emph{Euclidean Dynamical Triangulations} (DT) \cite{Ambjorn:1992aw}, can be understood as the gravitational counterpart of lattice field theories such as Lattice Quantum Chromodynamics \cite{Wilson:1974sk}, where instead of adding fields to a fixed lattice the dynamical lattice itself captures the gravitational degrees of freedom.

Focusing on the three-dimensional case, spacetime geometries in DT are constructed from gluing equilateral Euclidean tetrahedra along their bounding triangles into a piecewise flat metric on the 3-sphere.
Since the edge lengths are fixed, the only freedom is in the connectivity of the tetrahedra.
Computing the partition function of DT is essentially equivalent to the enumeration of 3-sphere triangulations with control on the number of tetrahedra and vertices.
Unfortunately, this combinatorial problem is still far beyond current mathematical reach, exemplified by the fact that it is still open whether the number of triangulations is bounded by an exponential in the number of tetrahedra (see \cite{budd2022family} for a recent discussion of the difficulties).
However, the combinatorial setup makes the model amenable to numerical study via the Markov Chain Monte Carlo technique.
The results of such studies \cite{Boulatov_phase_1991,agishtein1991three,Ambjoern_Three_1992,Ambjoern_vacuum_1992,Catterall_Entropy_1995,Hagura_Phases_1998,Hotta_Multicanonical_1998,Thorleifsson:1998nb} indicate that standard DT features a first-order phase transition between two non-physical phases: a \emph{branched polymer phase} in which the tetrahedra organize in an extended tree-like fashion and a \emph{crumpled phase} in which the triangulation is highly connected.

Various extensions of the model have been investigated in the hopes of observing new phases and/or continuous phase transitions, where critical phenomena may give rise to self-similar scaling limits of truly three-dimensional geometry.
One option is to adapt the model by introducing new contributions to the action, such as a non-trivial measure term (see e.g.\ \cite{bruegmann19934d,bilke19984dgauge,bilke19984d,Coumbe:2014nea} for studies in 4-dimensional DT), or by including matter fields on top of the triangulation \cite{AMBJORN1992253,ambjorn1993three,ambjorn2000recursive}.
Alternatively, one can restrict the class of triangulations used, such as in Causal Dynamical Triangulations \cite{ambjorn2001nonperturbative}, in which the triangulations admit a foliation by triangulated 2-spheres.
Whereas in some cases qualitatively different phases have been observed, critical phenomena for the random 3-sphere triangulations have not yet been observed in these numerical studies.

This work presents a renewed attempt, inspired by a recent combinatorial study \cite{budd2022family} as well as a numerical study \cite{buddcastro}.
The model of \cite{budd2022family} can be viewed as a hybrid of the above-mentioned adaptations: it involves both the introduction of additional degrees of freedom in the form of a pair of spanning trees, which one might attribute a matter interpretation, and a restriction of the class of triangulations.
We refer to triangulations of this type as \emph{triple-tree triangulations}, a precise definition of which follows below.
They satisfy several desirable properties: the spanning trees exhibit a local construction of the triangulation, which easily verifies its spherical topology and ensures exponential bounds on the enumeration, and they are shown to be in bijection with certain triples of plane trees.
Moreover, a preliminary numerical study indicates that in addition to a familiar branched polymer phase, the model exhibits another phase that is qualitatively different from the crumpled phase.
To corroborate these findings and connect them with the familiar phase diagram of DT, we consider a model with two extra bare coupling constants that allows one to interpolate between triple-tree triangulations and DT decorated by a pair of spanning trees, but without restrictions on the 3-sphere triangulations.
This search for new critical phenomena is supported by the numerical investigation of \cite{buddcastro}, where metric spaces constructed from triples of trees, although with a simpler distribution than the aforementioned triples, showed good scaling.

\section{Tree-decorated Dynamical Triangulations}

\subsection{Euclidean Dynamical Triangulations in 3D}

The problem of Euclidean quantum gravity in three dimensions can be formulated as the search for a non-perturbative construction of the formal Euclidean gravitational path integral
\begin{equation}
    \mathcal{Z} = \int \mathcal{D}[g_{\mu\nu}] e^{-S_{\mathrm{EH}}[g_{\mu\nu}]},
\end{equation}
where $S_{\mathrm{EH}}$ is the Einstein-Hilbert action and the integral should be over diffeomorphism equivalence classes of Riemannian metrics $g_{\mu\nu}$ on a 3-manifold, which we take here to be the 3-sphere. 
DT aims to construct this path integral via a continuum limit of a lattice discretization, in which the formally infinite-dimensional collection of Riemannian metrics is replaced by the countable subset of such metrics that can be obtained by gluing finitely many equilateral tetrahedra.
On these piecewise flat metrics the Einstein-Hilbert action reduces to a purely combinatorial expression \cite{Ambjoern_Three_1992,Ambjorn:2012jv}, involving the number of vertices $N_0$ and the number of tetrahedra $N_3$, with their corresponding bare coupling constant parameters, 
\begin{equation}
    S_{\mathrm{DT}} = -\kappa_0 N_0 + \kappa_3 N_3,
\end{equation}
where $\kappa_0$ is the bare gravitational coupling constant and $\kappa_3$ the bare cosmological constant.
The corresponding grand-canonical partition function
\begin{equation}
    \mathcal{Z}_{\mathrm{DT}}(\kappa_0,\kappa_3) = \sum_{\mathcal{T}\in \mathbb{T}} \frac{1}{C_\mathcal{T}}e^{-S_{\mathrm{DT}}[\mathcal{T}]}
    \label{eq:action}
\end{equation}
should be thought of as a discretization of the aforementioned path integral.
In $\mathcal{Z}_{\mathrm{DT}}(\kappa_0,\kappa_3)$ the sum is over a set $\mathbb{T}$ of (unlabeled) triangulations of the 3-sphere and $C_{\mathcal{T}}$ is the order of the automorphism group of $\mathcal{T}$.
As alluded to above, it is at present unknown whether $\mathcal{Z}_{\mathrm{DT}}(\kappa_0,\kappa_3) < \infty$ for any finite value of the couplings $\kappa_0$ and $\kappa_3$.
It therefore makes sense to focus on the canonical partition function
\begin{equation}
    \mathcal{Z}^{\mathrm{DT}}_N(\kappa_0) = \sum_{\mathcal{T}\in\mathbb{T}_N} \frac{1}{C_\mathcal{T}}e^{\kappa_0 N_0}, \qquad \mathcal{Z}_{\mathrm{DT}}(\kappa_0,\kappa_3) = \sum_{N=1}^\infty \mathcal{Z}^{\mathrm{DT}}_N(\kappa_0) e^{-\kappa_3 N},
\end{equation}
where $\mathbb{T}_N\subset \mathbb{T}$ is the set of triangulations with $N_3=N$ tetrahedra.

To unambiguously define the model, we should specify the precise class of 3-sphere triangulations used.
Here we follow the naming convention of \cite{Thorleifsson:1998nb}.
The most restrictive class is that of \emph{combinatorial} triangulations, in which edges connect distinct pairs of vertices and each edge or triangle is uniquely determined by its set of vertices.
In other words, combinatorial triangulations correspond precisely to simplicial complexes with 3-sphere topology.
These restrictions are lifted in the least restrictive class of \emph{degenerate} triangulations, which for instance allows an edge to start and end on the same vertex and a tetrahedron to be glued to itself by identifying two of its triangles.
Mathematically these correspond to pure CW-complexes with 3-sphere topology, in which all cells are simplices.
An intermediate class is that of \emph{restricted degenerate} triangulations, which are degenerate triangulations in which edges are required to connect distinct pairs of vertices.
In this case, all vertices of a triangle or a tetrahedron are distinct, but edges, triangles and tetrahedra are not necessarily uniquely determined by their sets of vertices.
In particular, tetrahedra cannot be glued to themselves, but a pair of tetrahedra may share more than one triangle.
In this work we will focus exclusively on the latter class and take $\mathbb{T}$ to be the set of restricted degenerate triangulations.

The reasons for this choice are mainly pragmatic.
First of all, having fewer restrictions can simplify the Markov chain used in the simulations, because local updates of the triangulation can be chosen to affect fewer tetrahedra at a time.
This is a particularly important consideration in our case because of the additional decoration that will be added in the form of spanning trees.
On the other hand, the class of degenerate triangulations has some features that lead to numerical instability in simulations, as already noted in \cite{Thorleifsson:1998nb}.
In particular, the vertex number in a degenerate triangulation with a fixed number of tetrahedra is maximized by a unique triangulation consisting of a long cyclic string of tetrahedra, each of which is glued to itself along a pair of its triangles.
This entails that at large $\kappa_0$, the typical branched polymer structure of DT degenerates into a deterministic configuration.
We have observed that at other values of $\kappa_0$ forbidding self-gluing of tetrahedra also improves numerical stability.

In the context of standard DT it is not really meaningful to ask whether this choice affects the continuum physics, since no universality class beyond that of the branched polymer has been identified.
The latter appears to emerge universally in simulations for each of the choices (provided $\kappa_0$ is not taken to its extreme for degenerate triangulations).
Only once an adaptation of the model gives rise to new critical phenomena, one may naturally ask whether they are universal with respect to the local combinatorial details of the model. 
This is not at all a given, because, in contrast to statistical systems living on regular lattices, the presence of high connectivity in random triangulations, like in the crumpled phase of DT, could cause local combinatorial restrictions to affect geometry globally.
In the case of two-dimensional DT, however, it is known that the continuum limit is independent of the class of triangulations \cite{legall2013,addario-berry2017} and universality holds much more generally. 
Previous studies in three-dimensional EDT and CDT have shown mostly quantitative differences \cite{Thorleifsson:1998nb,Brunekreef:2022pns}, where it is argued that the choice mainly affects the required size of triangulations needed to observe scaling behavior, typically favoring relaxed constraints.

\subsection{Spanning trees and middle graph}

In order to describe the adapted model, we need to recall some mathematical concepts.
A \emph{spanning tree} of a graph\footnote{More precisely we are considering multigraphs everywhere in this work, meaning that we allow multiple edges between pairs of vertices (as well as edges starting and ending at the same vertex, but those won't be relevant in the current setting).} is a connected acyclic subgraph containing all its vertices. To a triangulation $\mathcal{T}$ one can associate two natural graphs: its \emph{1-skeleton} or \emph{vertex graph} consisting of the vertices and edges of $\mathcal{T}$, and its \emph{dual graph} or \emph{tetrahedron graph} in which the tetrahedra and triangles of $\mathcal{T}$ respectively take the role of graph vertices and edges. Therefore there are also two natural notions of spanning trees of $\mathcal{T}$: \emph{vertex trees} and \emph{tetrahedron trees}, which are spanning trees of the respective graphs. 
We denote the sets of vertex trees and tetrahedron trees of $\mathcal{T}$ by $\mathbb{S}_\mathcal{T}$ and $\mathbb{S}^*_\mathcal{T}$ respectively.

\begin{figure}[ht]
    \centering
    \includegraphics[width = \textwidth]{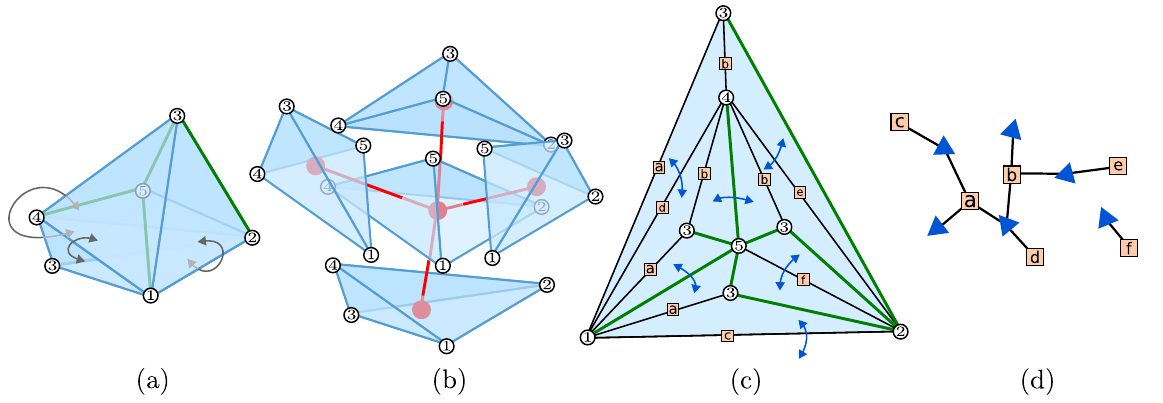}
    \caption{An example of a spanning-tree-decorated triangulation $(\mathcal{T},\mathcal{S},\mathcal{S}^*)$ with $N_3 = 5$ tetrahedra and $N_0 = 5$ vertices (labeled $1,\ldots,5$ in the figures). (a) The triangulation $\mathcal{T}$ with vertex tree $\mathcal{S}$ marked in green. (b) An exploded view of the tetrahedron tree $\mathcal{S}^*$. Vertices with the same label are identified in $\mathcal{T}$. (c) The boundary of the tetrahedron tree is a planar Apollonian triangulation with the edges associated to the vertex tree indicated in green. After identification of the triangles by the blue arrows, theses triangles and black edges with orange labels make up the vertex sets of the middle graph. (d) The middle graph $\mathcal{G}(\mathcal{T},\mathcal{S},\mathcal{S}^*)$ which has $N_{\mathrm{C}} = 2$ connected components and $N_{\mathrm{L}}=0$ loops.}
    \label{fig:tree-tetrahedra}
\end{figure}

To a triangulation $\mathcal{T}$ together with a distinguished vertex tree $\mathcal{S}\in \mathbb{S}_\mathcal{T}$ and tetrahedron tree $\mathcal{S}^*\in \mathbb{S}^*_\mathcal{T}$, one can associate another graph $\mathcal{G}(\mathcal{T},\mathcal{S},\mathcal{S}^*)$ called the \emph{middle graph} (see Fig.~\ref{fig:tree-tetrahedra} for an example). It is a bipartite graph whose vertex sets are the set of edges of $\mathcal{T}$ that are not part of $\mathcal{S}$ and the set of triangles not in $\mathcal{S}^*$, and an edge-triangle pair is adjacent in $\mathcal{G}(\mathcal{T},\mathcal{S},\mathcal{S}^*)$ if the edge is a side of the triangle in $\mathcal{T}$. 
It is a simple graph (i.e.\ it has at most one edge between a pair of vertices), because edges in a restricted degenerate triangulation cannot be adjacent to multiple sides of a triangle.
The middle graph is not necessarily connected, but it cannot have any isolated vertices.
This follows from the fact that not all triangles around an edge of the triangulation can be in the tetrahedron tree because that would amount to a cycle in the tree, nor can all three edges around a triangle be in the vertex tree.

Let us look at some natural invariants associated to the middle graph.
If $\mathcal{T}$ has $N_3$ tetrahedra and $N_0$ vertices, then Euler's relation implies that the number $N_1$ of edges and number $N_2$ of triangles of $\mathcal{T}$ are given by
\begin{equation}
    N_1 = N_0 + N_3, \qquad N_2 = 2 N_3.
\end{equation}
Since $\mathcal{S}$ and $\mathcal{S}^*$ are spanning trees, they traverse precisely $N_0-1$ edges and $N_3-1$ triangles respectively.
It follows that precisely $N_1 - (N_0-1) = N_3+1$ edges of $\mathcal{T}$ and $N_2 - (N_3-1) = N_3+1$ triangles of $\mathcal{T}$ make up the vertex sets of the middle graph $\mathcal{G}(\mathcal{T},\mathcal{S},\mathcal{S}^*)$, so the total number of vertices is $N_\mathrm{v} = 2N_3+2$.
Besides the number of vertices, there are three more natural graph invariants: the number $N_{\mathrm{e}}$ of edges, the \emph{circuit rank} or \emph{loop number} $N_{\mathrm{L}}$ (i.e.\ the minimal number of edges that need to be removed to make the graph acyclic), and the number $N_{\mathrm{C}}$ of connected components.
These are well-known to satisfy the linear relation
\begin{equation}\label{eq:middlegraphrelation}
    N_{\mathrm{e}} = N_{\mathrm{v}} - N_{\mathrm{C}} + N_{\mathrm{L}},
\end{equation}
leaving us with two independent invariants, which we take to be $N_{\mathrm{C}}$ and $N_{\mathrm{L}}$.
Indeed, one can show that no non-trivial linear combination of these is determined by $N_0$ and $N_3$.

With this in mind we may naturally introduce a statistical system of random pairs of spanning trees on a triangulation $\mathcal{T}$ by sampling $(\mathcal{S},\mathcal{S}^*)$ with probability
\begin{equation}
\frac{e^{\beta_{\mathrm{L}}N_{\mathrm{L}}+\beta_{\mathrm{C}}(N_{\mathrm{C}}-1)}}{\mathcal{Z}_{\mathcal{T}}(\beta_{\mathrm{L}},\beta_{\mathrm{C}})}, \qquad \mathcal{Z}_{\mathcal{T}}(\beta_{\mathrm{L}},\beta_{\mathrm{C}}) = \sum_{\mathcal{S}\in\mathbb{S}_{\mathcal{T}}}\sum_{\mathcal{S}^*\in\mathbb{S}^*_{\mathcal{T}}}e^{\beta_{\mathrm{L}}N_{\mathrm{L}}+\beta_{\mathrm{C}}(N_{\mathrm{C}}-1)},
\end{equation}
where $\beta_{\mathrm{L}}$ and $\beta_{\mathrm{C}}$ are bare coupling constants that introduce interactions between $\mathcal{S}$ and $\mathcal{S}^*$.
When $\beta_{\mathrm{L}}=\beta_{\mathrm{C}}=0$, the trees are independent of each other and distributed as the \emph{uniform spanning trees} on the vertex graph and tetrahedron graph respectively (see e.g.\ \cite{pemantle2004uniform} for an introduction to uniform spanning trees).

\subsection{Model definition: tree-decorated dynamical triangulations}\label{sec:model}

Coupling this system to dynamical triangulations leads to the canonical partition function of tree-decorated dynamical triangulations, that is at the center of this work,
\begin{align}
    \mathcal{Z}_N(\kappa_0,\beta_{\mathrm{L}},\beta_{\mathrm{C}}) &= \sum_{\mathcal{T}\in \mathbb{T}_N} \frac{1}{C_\mathcal{T}}e^{\kappa_0N_0} \mathcal{Z}_{\mathcal{T}}(\beta_{\mathrm{L}},\beta_{\mathrm{C}}) = \sum_{\mathcal{T}\in \mathbb{T}_N} \frac{1}{C_\mathcal{T}} \sum_{\mathcal{S}\in\mathbb{S}_{\mathcal{T}}}\sum_{\mathcal{S}^*\in\mathbb{S}^*_{\mathcal{T}}} e^{-S_{\mathrm{DT}}^{\mathrm{mg}}}, \\
    S_{\mathrm{DT}}^{\mathrm{mg}}(\mathcal{T},\mathcal{S},\mathcal{S}^*) &= -\kappa_0 N_0 -\beta_{\mathrm{L}} N_{\mathrm{L}} - \beta_{\mathrm{C}} (N_{\mathrm{C}}-1).
\end{align}
The grand-canonical partition function $\mathcal{Z}(\kappa_0,\kappa_3,\beta_{\mathrm{L}},\beta_{\mathrm{C}})$ is defined accordingly.

To appreciate some aspects of the model, let us focus on several special values of $\kappa_0$, $\beta_{\mathrm{L}}$ and $\beta_{\mathrm{C}}$.
First, the case $\beta_{\mathrm{L}}=\beta_{\mathrm{C}}=0$, for which $\mathcal{Z}_{\mathcal{T}}(0,0)$ simply counts the number of pairs consisting of a vertex tree and a tetrahedron tree.
Denoting by $\Delta$ and $\Delta^*$ the graph laplacian of the vertex graph and tetrahedron graph respectively, we know by Kirchhoff's matrix tree theorem that
\begin{equation}
    \mathcal{Z}_{\mathcal{T}}(0,0) = {\det}'(\Delta) \,{\det}'(\Delta^*), 
\end{equation}
where $\det'(\Delta)$ denotes the cofactor of the matrix $\Delta$, meaning the determinant of $\Delta$ after removing the $i$th column and $i$th row for any choice of $i$.
Therefore, coupling to random triangulations yields a reweighted version of DT,
\begin{equation}
    \mathcal{Z}_N(\kappa_0,0,0) = \sum_{\mathcal{T}\in \mathbb{T}_N} \frac{{\det}'(\Delta) \,{\det}'(\Delta^*)}{C_\mathcal{T}}.
\end{equation}
The same partition function can be obtained by coupling DT to a pair of free fermionic fields, one living on the vertices and one on the tetrahedra of the triangulation with nearest neighbour interactions. 
One should therefore view the case $\beta_{\mathrm{L}}=\beta_{\mathrm{C}}=0$ as a mild modification of standard DT.
In particular, preliminary simulations indicated that this modification does not affect the phase structure qualitatively: at large $\kappa_0$ one finds a branched polymer phase, separated by a first-order phase transition from a crumpled phase at small $\kappa_0$.  

It is worth noting that in the case of two-dimensional DT, coupling to a uniform spanning tree or, equivalently, reweighting by the cofactor of the vertex laplacian, changes the universality class to that of gravity coupled to a conformal field theory with central charge $c=-2$ \cite{mullin1967,kazakov1985critical,BOULATOV1986641}.
The main reason this model has received significant attention is because of its bijective correspondence with pairs of plain trees, originally due to Mullin \cite{mullin1967,bernardi2007}, which also provides an efficient way of sampling random spanning-tree decorated triangulations in Monte Carlo studies \cite{Kawamoto:1991xd,Ambjorn:1996kb,Ambjorn:2011rs}.
In recent years it has become clear that this encoding is a key example of the mating of trees framework \cite{duplantier2021} that applies to two-dimensional quantum gravity coupled to more general types of critical matter systems.

Mullin's bijection relies on the fact that a choice of a spanning tree on the vertex graph of a two-dimensional triangulation determines a second spanning tree on the dual graph (i.e.\ the triangle graph).
Vice versa, these two trees viewed properly as plane trees fully determine the triangulation.
This is not the case for triangulations of the 3-sphere: the tetrahedron tree can be chosen independently of the vertex tree, and the additional structure of the middle graph is necessary to uniquely determine the three-dimensional triangulation.
This brings us to examine the case $\beta_{\mathrm{L}} = \beta_{\mathrm{C}} = -\infty$, such that $\mathcal{Z}_{\mathcal{T}}(-\infty,-\infty)$ enumerates the pairs $(\mathcal{S},\mathcal{S}^*)$ of trees for which the middle graph $\mathcal{G}(\mathcal{T},\mathcal{S},\mathcal{S}^*)$ is a tree as well, meaning that it is acyclic ($N_{\mathrm{L}}=0$) and connected ($N_{\mathrm{C}}=1$).
The corresponding canonical partition function $\mathcal{Z}_{N}(\kappa_0,-\infty,-\infty)$ thus involves a sum over 3-sphere triangulations decorated with a triple of trees.
In this case one can meaningfully ask whether knowledge of the three trees, encoded in an appropriate combinatorial fashion, determines the triangulation, in analogy with the two-dimensional case.
This is precisely what was established in \cite{budd2022family}, as we shall now summarize.

\begin{figure}[ht]
    \centering
    \includegraphics[width = \textwidth]{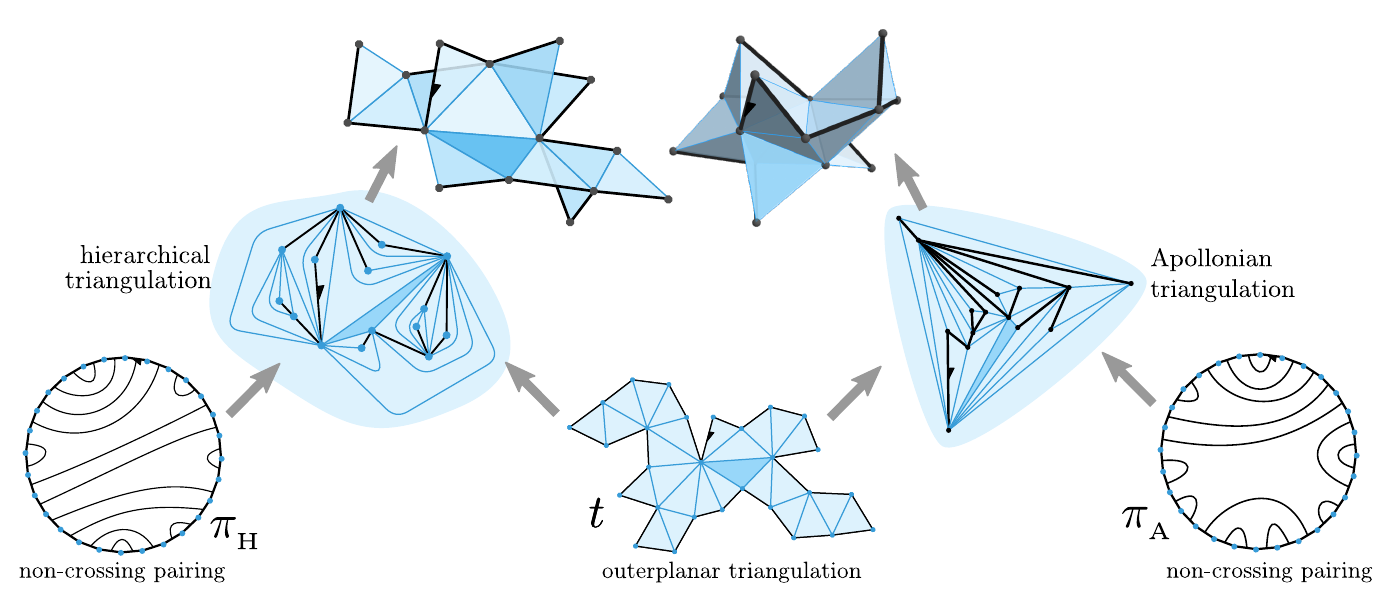}
    \caption{Illustration of triple-tree bijection of \cite{budd2022family}: a triple-tree consists of an outerplanar triangulation $t$ of an $n$-gon and a pair of non-crossing pairings $\pi_{\mathrm{H}}, \pi_{\mathrm{A}}$ of the $n$-gon, with the restriction that gluing the boundary of $t$ according to $\pi_{\mathrm{H}}$ (resp.\ $\pi_{\mathrm{A}}$) yields a hierarchical triangulation (resp.\ Apollonian triangulation) of the 2-sphere. Figure adapted from \cite{budd2022family}.}
    \label{fig:hierarchicalapollonian}
\end{figure}

The appropriate fashion to describe the three trees turns out to be via a \emph{triple-tree} $(t,\pi_{\mathrm{H}},\pi_{\mathrm{A}})$.
Here $t$ is an \emph{outerplanar triangulation} of the $(2N+4)$-gon, i.e.\ a triangulation of an $(2N+4)$-sided polygon with no vertices in the interior of the polygon, and $\pi_{\mathrm{H}},\pi_{\mathrm{A}}$ both are \emph{non-crossing pairings} of the $(2N+4)$-gon, i.e.\ pairings of the sides of the polygon that can be drawn by non-intersecting arcs in its interior (see Fig.~\ref{fig:hierarchicalapollonian}).
All three can be viewed dually as plane trees, justifying the terminology triple-tree.
Several restrictions need to be imposed on the triple-tree.
When the boundary sides of $t$ are glued according to the pairing $\pi_{\mathrm{H}}$ the result is a triangulated 2-sphere that is required to be a so-called \emph{hierarchical triangulation}, a triangulation that can be obtained from a single triangle by repeatedly replacing an edge by a pair of triangles glued along two sides.
Similarly, gluing $t$ according to $\pi_{\mathrm{A}}$ should give an \emph{Apollonian triangulation}, a triangulation that can be obtained from a single triangle by repeatedly subdividing a triangle into three triangles with a new vertex in the center.
According to \cite[Theorem~1]{budd2022family} such triple trees are in bijection with tree-decorated triangulations $(\mathcal{T},\mathcal{S},\mathcal{S}^*)$ with $N$ tetrahedra such that the middle graph is a tree (with an appropriate choice of rooting that takes care of the symmetry factor $1/C_{\mathcal{T}}$).
We should note, however, that in \cite{budd2022family} the triangulation $\mathcal{T}$ was assumed more generally to be of the degenerate class.
To get only the restricted degenerate triangulations $\mathcal{T}$ in our setting, one should impose a further restriction on $(t,\pi_{\mathrm{H}},\pi_{\mathrm{A}})$, namely that the identification of the boundary sides of $t$ under $\pi_{\mathrm{H}}$ and $\pi_{\mathrm{A}}$ combined does not result in an identification of the endpoints of any of the diagonals of $t$.
Hence, $\mathcal{Z}_N(\kappa_0,-\infty,-\infty)$ can equivalently be understood as the partition function of triple trees $(t,\pi_{\mathrm{H}},\pi_{\mathrm{A}})$ with a Boltzmann weight involving $\kappa_0$ that depends on $\pi_{\mathrm{H}}$ and $\pi_{\mathrm{A}}$ in a simple manner (see \cite{budd2022family} for details).
It is easily seen that there are at most exponentially many triple-trees, implying that the grand-canonical partition function $\mathcal{Z}(\kappa_0,\kappa_3,-\infty,-\infty)$ is convergent for sufficiently large $\kappa_3$.
Moreover, preliminary Monte Carlo simulations of these triple-trees pointed at the existence of a phase transition between a branched polymer phase at large $\kappa_0$ and a new phase, referred to as the \emph{triple-tree phase}, at small $\kappa_0$.
Since our model allows to interpolate between the unweighted tree-decorated DT (at $\beta_{\mathrm{L}} = \beta_{\mathrm{L}} = 0$) and the triple-tree class (at $\beta_{\mathrm{L}} = \beta_{\mathrm{L}} = -\infty$), one can study how their phase transitions fit into the extended $3$-dimensional phase diagram parametrized by $\kappa_0$, $\beta_{\mathrm{L}}$ and $\beta_{\mathrm{C}}$.

Further extremes we may consider are $\kappa_0 =\pm\infty$, which restricts the triangulations to have minimal or maximal number of vertices for fixed number $N_3=N$ of tetrahedra.
Due to the restricted degenerate conditions, the vertices of tetrahedra are required to be distinct.
Therefore the minimal number of vertices $N_0=4$ is achieved when all tetrahedra share the same 4 vertices.
Such triangulations are dual to certain colored graphs known as \emph{crystallizations} of the 3-sphere in the literature \cite{ferri1986graph}.
Therefore, $\lim_{\kappa_0\to-\infty} e^{-4\kappa_0}\mathcal{Z}_N(\kappa_0,\beta_{\mathrm{L}},\beta_{\mathrm{C}})$ can be understood as a partition function of tree-decorated crystallizations.

The maximal number of vertices $N_0 = \frac{N_3}{2}+3$ for $N_3=N$ even is achieved by the \emph{melonic triangulations} \cite{bonzom2011critical} of the 3-sphere.
A proof of this fact is given in Appendix~\ref{sec:melonicbound}.
These can be obtained from the minimal triangulation, consisting of two tetrahedra sharing all four triangles, by repeatedly replacing a selected triangle by a pair of tetrahedra that are glued along three of their triangles (see the triangular pillow move below). 
Therefore $\lim_{\kappa_0\to\infty} e^{-(3+N/2)\kappa_0}\mathcal{Z}_N(\kappa_0,\beta_{\mathrm{L}},\beta_{\mathrm{C}})$ is the partition function of tree-decorated melonic triangulations.

Finally, one could consider regimes with large $\beta_{\mathrm{L}}$ or $\beta_{\mathrm{C}}$.
For this it is relevant to know the following upper bounds,
\begin{equation}
    N_{\mathrm{C}} \leq N_3 + 1, \qquad N_{\mathrm{L}} + N_0 \leq N_3 + 3.
\end{equation}
The first of these follows from the earlier observations that the middle graph has $N_{\mathrm{v}} = 2N_3+2$ vertices and that each connected component contains at least two vertices.
The proof of the second bound is given in Appendix~\ref{sec:NLbound}.

\section{Monte Carlo methods and implementation}

We wish to understand the phase structure of the partition function $\mathcal{Z}_N(\kappa_0,\beta_{\mathrm{L}},\beta_{\mathrm{C}})$ in the large-$N$ limit.
Our numerical study relies on Markov chain Monte Carlo methods to sample random triangulations from the corresponding Boltzmann distribution at fixed finite $N$ to estimate expectation values of observables, after which their scaling behaviour with increasing $N$ can be analyzed.
Such Monte Carlo methods have been extensively applied in DT (see \cite{Ambjorn:2012jv} for an overview), but tree decoration presents several challenges.
For this reason, we describe the algorithm and implementation in some detail.

\subsection{Markov chain}

Although we are interested in sampling from the canonical partition function $\mathcal{Z}_N(\kappa_0,\beta_{\mathrm{L}},\beta_{\mathrm{C}})$, devising an ergodic Markov chain on the ensemble of 3-sphere triangulation with a fixed number $N_3=N$ of tetrahedra is hard.
As is customary in DT simulations, instead one constructs an ergodic Markov chain on triangulations of arbitrary size in such a way that in the stationary distribution the random number of tetrahedra $N_3$ is concentrated around $N$ and such that the marginal distribution conditionally on $N_3=N$ agrees with that of $\mathcal{Z}_N(\kappa_0,\beta_{\mathrm{L}},\beta_{\mathrm{C}})$.
This can be achieved by introducing a volume-fixing potential in the grand-canonical partition function,
\begin{align}
    \mathcal{Z}_{\varepsilon,N}(\kappa_0,\kappa_3,\beta_{\mathrm{L}},\beta_{\mathrm{C}}) &= \sum_{N_3 = 1}^\infty e^{-\kappa_3 N_3 - \varepsilon(N_3-N)^2} \mathcal{Z}_{N_3}(\kappa_0,\beta_{\mathrm{L}},\beta_{\mathrm{C}})\nonumber\\
    &= \sum_{\mathcal{T}\in \mathbb{T}_N} \frac{1}{C_\mathcal{T}} \sum_{\mathcal{S}\in\mathbb{S}_{\mathcal{T}}}\sum_{\mathcal{S}^*\in\mathbb{S}^*_{\mathcal{T}}} e^{-S_{\varepsilon,N}}, \label{eq:gcensemblevolumefixing}\\
    S_{\varepsilon,N}(\mathcal{T},\mathcal{S},\mathcal{S}^*) &= \kappa_3 N_3 + \varepsilon(N_3-N)^2-\kappa_0 N_0 -\beta_{\mathrm{L}} N_{\mathrm{L}} - \beta_{\mathrm{C}} (N_{\mathrm{C}}-1).\label{eq:actionvolumefixing}
\end{align}
The values of $\kappa_3$ and $\varepsilon$ are then tuned to ensure sufficiently small variance for $N_3$ while $\langle N_3 \rangle \approx N$.
Once we have an equilibrated Markov chain with this stationary distribution, discarding any states with $N_3 \neq N$ will leave us with samples distributed according to $\mathcal{Z}_N(\kappa_0,\beta_{\mathrm{L}},\beta_{\mathrm{C}})$.

If we disregard the presence of the trees, it is well-known how to construct an ergodic Markov chain on triangulations of the 3-sphere.
Depending on the precise combinatorial restrictions on the triangulations, any two such triangulations are connected by a finite number of local Pachner moves or variations thereof.
In the case of restricted degenerate triangulations, it is sufficient to consider two such local moves (and their inverses): the \emph{bistellar flip} move, that replaces two adjacent tetrahedra by a triple of tetrahedra sharing a single new edge in the triangulation, and the \emph{triangular pillow} move (also known as \emph{3-dipole} move), that inflates a triangle by inserting a pair of tetrahedra glued along three of their faces.
It turns out to be beneficial for the mixing of the Markov chain, especially in the regime of negative $\kappa_0$ where triangulations have few vertices, to supplement these with the \emph{quadrangular pillow} move (also known as \emph{2-dipole} move), that inflates a pair of triangles sharing an edge by inserting a pair of tetrahedra glued along two of their faces. 
A Markov chain with desired stationary distribution can then be constructed by applying the Metropolis-Hastings algorithm to a proposal transition matrix based on random selection of one of these moves.

\begin{figure}[ht]
    \centering
    \includegraphics[width = \textwidth]{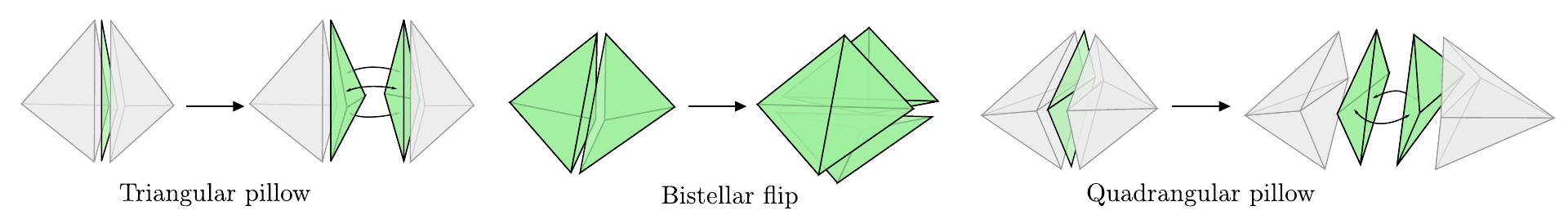}
    \caption{The local moves on the triangulation, which supplemented by their inverses give 6 types of local moves on the triangulation.}
    \label{fig:geometricmoves}
\end{figure}

In our case the state space consists of triples $(\mathcal{T},\mathcal{S},\mathcal{S}^*)$ consisting of triangulations decorated with a pair of trees. 
Therefore, several complications must be taken into account.
First of all, we should make sure that the moves of the Markov chain produce properly decorated triangulations.
Secondly, we should make sure that the moves are ergodic on this larger state space.
Thirdly, the Metropolis-Hastings algorithm should take into account the Boltzmann weights associated to the middle graph $\mathcal{G}(\mathcal{T},\mathcal{S},\mathcal{S}^*)$ in order to ensure convergence to the distribution of \eqref{eq:gcensemblevolumefixing}.
To ensure ergodicity, two new moves are introduced that change only the tree $\mathcal{S}$ or $\mathcal{S}^*$.
Below we briefly discuss each of the moves, before explaining their ergodicity.

\subsubsection{Triangular pillow move}
The triangular pillow move selects a uniform random triangle $t$ in $\mathcal{T}$ and proposes to inflate the triangle $t$ into a pair of tetrahedra glued along three of its sides.
Since this adds a new vertex to the triangulation, the vertex tree $\mathcal{S}$ is supplemented with one of the three new edges chosen uniformly at random.
Depending on whether $t$ belongs to $\mathcal{S}^*$ or not, there are 3 respectively 7 ways to adapt $\mathcal{S}^*$ into a tetrahedron tree of the resulting triangulation, of which again one is chosen uniformly at random.

For the inverse triangular pillow move, we chose a uniform random vertex $v$ of coordination number 2, meaning that it is incident to precisely 2 tetrahedra.
If more than one of the three edges incident to $v$ is in $\mathcal{S}$, the move must be rejected because such a configuration cannot be the result of a triangular pillow move.
Otherwise, the deletion of the two adjacent tetrahedra can be applied and there is a unique way to adapt $\mathcal{S}$ and $\mathcal{S}^*$ into spanning trees of the resulting triangulation.

To ensure detailed balance for the triangular pillow and its inverse, when choosing between the two with a fair coin flip, the appropriate Metropolis-Hastings acceptance probability for the move becomes\footnote{The simulation actually uses a multi-stage acceptance/rejection for most moves, which effectively results in a lower acceptance probability than stated here but one that still satisfies the detailed balance condition. This is done because it is computationally beneficial to reject early based on computationally cheap tests, while minimizing the rejection rate after computationally expensive tests. For instance, computing the change in $N_3$ and $N_0$ is very efficient while computing the change in $N_{\mathrm{L}}$ and $N_{\mathrm{C}}$ is much more time consuming.}
\begin{equation}
    \mathcal{A}_{\text{triang.\ pillow}} = \min\left(1,\frac{3f \cdot 2N_3}{C^{\mathrm{V}}_2}e^{-\Delta S}\right),\qquad f = \begin{cases}
        3 & t\in \mathcal{S}^*\\
        7 & t\notin \mathcal{S}^*,
    \end{cases}
\end{equation}
where $C^{\mathrm{V}}_2$ is the number of vertices of coordination number 2 in the triangulation after the move and $\Delta S$ is the change in the action \eqref{eq:actionvolumefixing} caused by the move.
The acceptance probability for the inverse move is
\begin{equation}
    \mathcal{A}_{\text{inv.\ triang.\ pillow}} = \min\left(1,\frac{C^{\mathrm{V}}_2}{3 f \cdot 2(N_3 - 2)}e^{-\Delta S}\right),
\end{equation}
where $C^{\mathrm{V}}_2$ is the number of vertices of coordination number 2 before the move and $f$ is as before depending on the tetrahedron tree after the move.
Note that in both cases the middle graph changes and therefore $\Delta S$ contains a contribution for the change in the number of loops $N_{\mathrm{L}}$ and number of connected components $N_{\mathrm{C}}$.

\subsubsection{Bistellar flip move}

The bistellar flip move selects a uniform random triangle $t$ in $\mathcal{T}$.
If the vertices $v_1,v_2$ opposite to $t$ in the two neighbouring tetrahedra are distinct, the move creates a new edge $e$ connecting $v_1$ and $v_2$ and replaces the two tetrahedra by three tetrahedra organized around $e$.
Since this changes the vertex graph only by the addition of the edge $e$, the vertex tree $\mathcal{S}$ is not changed (i.e.\ the edge is not included in the new vertex tree).

The changes in the tetrahedron tree are more complicated, because before and after the move there can be many local configurations of $\mathcal{S}^*$ in the two respectively three tetrahedra involved, which can be encoded by specifying for each of the 7 respectively 9 involved triangles whether they are in the tree. 
We therefore make use of a hard-coded lookup table, in which such local configuration before the move are related to a number of potential configurations after the move, from which one is chosen uniformly at random (or the move is rejected if there are none).
The local configurations, however, do not fully ensure that the update leads to a valid tetrahedron tree globally, so this needs to be verified and the move rejected in case of failure.

For the inverse move a uniform random edge $e$ of coordination number 3 is selected. 
If $e \in \mathcal{S}$, the move has to be rejected.
Otherwise replacing the three tetrahedra around $e$ by a pair of tetrahedra, results in a valid restricted degenerate triangulation with unchanged vertex tree.
For the update in the tetrahedron tree the same lookup table as before is used.

The appropriate Metropolis-Hastings acceptance probabilities thus become
\begin{equation}
    \mathcal{A}_{\text{bistellar flip}} = \min\left(1,\frac{B_{1}}{B_2}\frac{2N_3}{C^{\mathrm{E}}_3}e^{-\Delta S}\right),
\end{equation}
where $B_1$ respectively $B_2$ is the number of times the initial respectively final local configuration of the tetrahedron tree appears in the lookup table, and $C^{\mathrm{E}}_3$ is the number of edges of coordination number 3 after the move.
Similarly, for the inverse move, 
\begin{equation}
    \mathcal{A}_{\text{inv.\ bistellar flip}} = \min\left(1,\frac{B_{1}}{B_2}\frac{C^{\mathrm{E}}_3}{2(N_3-1)}e^{-\Delta S}\right),
\end{equation}
where now $C^{\mathrm{E}}_3$ is the number of edges of coordination number 3 before the move.

\subsubsection{Quadrangular pillow move}

The quadrangular pillow move selects a uniform random triangle $t_1$ in $\mathcal{T}$ and uniformly one of its three bounding edges $e$.
Then a second triangle $t_2$ distinct from $t_1$ is chosen uniformly among the triangles incident to $e$.
If the vertices opposite $e$ in $t_1$ and $t_2$ are distinct, the move proposes inflating the rhombus formed by $t_1$ and $t_2$ by the insertion of two tetrahedra glued along two of their sides.
From the point of view of the vertex graph, the only change is that the edge $e$ is doubled and a new edge $e'$ is introduced.
The latter should not be included in the vertex tree.
In case $e \in \mathcal{S}$, one of the two copies of $e$ is chosen uniformly to remain in the vertex tree.
For the tetrahedron, one has to distinguish three cases.
If $t_1,t_2 \in \mathcal{S}^*$, the move is rejected because the tetrahedron tree cannot be appropriately updated.
If one of $t_1,t_2$ is in $\mathcal{S}^*$, there are $2$ options for the resulting tetrahedron tree, since it should include either one of the two triangles shared by the new tetrahedra.
If none of them are in $\mathcal{S}^*$, a total of 12 options can be identified.
In either case we choose an option uniformly at random.

For the inverse quadrangular pillow move a uniform random edge $e'$ of coordination number 2 is selected.
The move is rejected if $e' \in \mathcal{S}$, or if the two tetrahedra incident to $e$ share more than two of their triangles, or if the tetrahedron tree is not of the type that can be produced by the quadrangular pillow move.
Otherwise, the move proposes to delete $e'$ and the adjacent tetrahedra and to close the gap remaining in the triangulation.
The vertex tree and tetrahedron tree on the resulting triangulation are then uniquely determined.

Accordingly, the Metropolis-Hastings acceptance probabilities are
\begin{equation}
    \mathcal{A}_{\text{quad.\ pillow}} = \min\left(1,\frac{f_1f_2 \cdot 6N_3 (c_e-1)}{C^{\mathrm{E}}_2}e^{-\Delta S}\right),\quad f_1 = \begin{cases}
        2 & e \in \mathcal{S}\\
        1 & e \notin \mathcal{S},
    \end{cases} \quad f_2 = \begin{cases}
        12 & t_1,t_2 \notin \mathcal{S}^*\\
        2 & \text{otherwise},
    \end{cases}
\end{equation}
where $c_e$ is the coordination number of the edge $e$ and $C^{\mathrm{E}}_2$ is the number of edges of coordination number 2 in the triangulation after the move.
Similarly,
\begin{equation}
    \mathcal{A}_{\text{inv.\ quad.\ pillow}} = \min\left(1,\frac{C^{\mathrm{E}}_2}{f_1f_2 \cdot 6(N_3-2) (c_e-1)}e^{-\Delta S}\right),
\end{equation}
where now $C_2^{\mathrm{E}}$ is the number of edges of coordination number 2 in the triangulation before the move, while $f_1$, $f_2$ and $c_e$ are computed as above but in the triangulation after the move.

\subsubsection{Tree update moves}
The Markov chain also includes moves that update one of the trees, while leaving the other tree and the triangulation unchanged.
The move is an adaptation of the well-known edge-swapping Markov chain \cite{broder1989generating} on spanning trees of a general graph that has the uniform spanning tree as the stationary distribution.
In the case of the vertex tree, one of the $N_3+1$ edges of the $\mathcal{T}$ that is not in $\mathcal{S}$ is chosen uniformly at random. 
Denoting this edge by $e$, the graph $\mathcal{S} \cup \{e\}$ will have a unique cycle.
Removing a uniform random edge (not equal to $e$ itself) from this cycle yields a new spanning tree of the vertex graph.
To ensure detailed balance we only need to account for the changes in the number of loops and connected components in the middle graph in the Metropolis-Hastings acceptance probability
\begin{equation}
    \mathcal{A}_{\text{tree\ update}} = \min\left(1,e^{-\Delta S}\right).
\end{equation}
The algorithm for updating the tetrahedron tree $\mathcal{S}^*$ is completely analogous. 

\subsubsection{Ergodicity}

The full Markov chain on the state space of tree-decorated triangulations $(\mathcal{T},\mathcal{S},\mathcal{S}^*)$ consists in choosing at each step one of the 5 types (counting the vertex tree and tetrahedron tree update separately) of moves randomly with fixed distribution.
Since detailed balance was ensured for each move independently, this means the desired Boltzmann distribution \eqref{eq:gcensemblevolumefixing} is stationary.
It remains to check ergodicity of the Markov chain in order to guarantee convergence in distribution to its stationary distribution.

It is well known that any pair of 3-sphere triangulations can be connected with a finite sequence of Pachner moves, and similarly any pair of spanning trees can be connected via a finite sequence of tree updates.
The only issue is that some of the Pachner moves in our case are accepted with $0$ probability, because the particular configuration of trees $(\mathcal{S},\mathcal{S}^*)$ prohibits a move which would be valid from the geometric point of view.
Luckily, we only need to verify that for each triangulation $\mathcal{T}$ and choice of geometrically valid Pachner move, there exists at least one pair of trees $(\mathcal{S},\mathcal{S}^*)$ on $\mathcal{T}$ for which the move is accepted with positive probability.
This is because every tree update is accepted with positive probability.

The triangular pillow move does not pose any troubles, while the inverse triangular pillow move at vertex $v$ requires that $v$ is the leaf of the vertex tree $\mathcal{S}$.
Of course, such a vertex tree always exists.
For the bistellar flip move there are no restrictions on the vertex tree, while for its inverse one only needs to make sure that the edge $e$ is not in the vertex tree.
With regards to the tetrahedron tree $\mathcal{S}^*$ before the bistellar flip move, one may check that if the triangle $t \in \mathcal{S}^*$, then there always exists a valid tetrahedron tree after the move that includes two of the three triangles around the edge $e$.
And vice versa for the inverse bistellar flip move.
Therefore, ergodicity of the Markov chain is granted (at least if all the bare coupling constants take finite values).

\subsection{Implementation}

Having established a Markov chain that converges to the appropriate Boltzmann distribution, we could go ahead and represent the triple $(\mathcal{T},\mathcal{S},\mathcal{S}^*)$ in a simple data structure and apply the outlined moves.
The problem is that without careful choice of the data structure the computational complexity of a single move grows rapidly with the system size, because the requirement on $\mathcal{S}$ and $\mathcal{S}^*$ to remain spanning trees is not a local one.
The same is true for the determination of the invariants $N_{\mathrm{L}}$ and $N_{\mathrm{C}}$ of the middle graph required for the Metropolis-Hastings tests.
These can generally not be inferred locally from the structure of the triangulation and trees.
For instance, a change of the number $N_{\mathrm{C}}$ of connected components of the middle graph by the removal of a single edge may depend on the presence of another edge at the opposite end of the triangulation.

In order to achieve a polylogarithmic (amortized) time complexity $O(\log^2 N)$ per move, we make use of the following data structures.
The trees $\mathcal{S}$ and $\mathcal{S}^*$ are stored in a \emph{dynamic tree} or \emph{link/cut tree} data structure, introduced by Sleator and Tarjan \cite{sleator1981data}.
It allows for adding, removing and querying links in a forest (i.e. a collection of trees) in logarithmic time, which are needed for testing validity of tree updates for the Pachner moves and for updating the spanning trees.
Moreover, for the purpose of the tree update moves, it allows determining the length of the unique cycle and uniformly sampling an edge from it in logarithmic time as well.
Our simulation code relies on the convenient C++ implementation provided by David Eisenstat \cite{dtree}.

Even though the middle graph $\mathcal{G}(\mathcal{T},\mathcal{S},\mathcal{S}^*)$ is fully determined by the triangulations and trees, it needs to be encoded and updated during the simulation to achieve efficient computation of the change in its number  $N_{\mathrm{C}}$ of connected components.
Note that then changes in the loop number $N_{\mathrm{L}} = N_{\mathrm{e}} - N_{\mathrm{v}} + N_{\mathrm{C}}$ can be computed easily, because the number of edges $N_{\mathrm{e}}$ and vertices $N_{\mathrm{v}}$ of the middle graph are local quantities.
The problem of keeping track of connectivity information in a graph that gets updated by addition and removal of edges is known as \emph{fully dynamic connectivity}.
We rely on the data structure of Holm, de Lichtenberg and Thorup \cite{polylog}, which achieves amortized time complexity $O(\log^2 N)$ for updates and connectivity queries (and makes use of the dynamic tree data structure under the hood).
Our code makes use of the C++ implementation of Tom Tseng \cite{Tom-Tseng}.

The full C++ code used for the simulations in this work is available in the Github repository
\begin{center}
\url{https://github.com/Kregnach/Euclidean-Dynamical-Triangulations-with-Link-Cut-Trees-and-Dynamic-Connectivity}
\end{center}

\subsection{Data Analysis}

The main goal of this work is to explore the three-dimensional phase diagram of tree-decorated DT and to characterize the nature of the phase transitions present.
Since phase transitions only occur in the limit of infinite system size and we are limited to simulating finite systems, this necessarily entails a scaling analysis.
The first step involves a rough exploration of the phase diagram, searching for regions where expectation values of observables depend sensitively on the coupling constants.
Since performing simulations for a full three-dimensional grid of coupling constants would be computationally too expensive, we choose in this work to focus on extreme values of the coupling constants on the boundary of a box in parameter space.
Once the rough location of a potential phase transition is determined together with observables that can serve as order parameters for the transition, one can start a more detailed scaling analysis in its vicinity.
According to the Ehrenfest classification of phase transitions, a transition is of order $n$ if the $n$th derivative of the free energy as function of the coupling constants has a discontinuity\footnote{Not every phase transition can be characterized like that, for example, phases featuring long-range correlations or topological order are excluded}. In case of a continuous phase transition, the Landau paradigm predicts typical scaling characteristics of the system.
Although it is not a given that this framework applies to the gravitational setting without a fixed background geometry, it is standard to assume that general characteristics persist. 
One of these characteristics relates to the scaling of the susceptibility $\chi_\mathcal{O}(N_3)$ of an order parameter $\mathcal{O}$,
\begin{equation}
    \chi_\mathcal{O} = \frac{1}{N_3} \operatorname{Var}(\mathcal{O}) = \frac{1}{N_3}( \langle \mathcal{O}^2\rangle - \langle \mathcal{O}\rangle^2)
\end{equation}
where $N_3$ is the system size.
If we denote by $\beta$ the coupling constant that induces a transition witnessed by the order parameter $\mathcal{O}$, then generally $\chi_\mathcal{O}(\beta)$ displays a peak for finite values of $N_3$ that becomes sharper as $N_3\to\infty$.
Two critical exponents, $\nu$ and $\gamma$, can be associated to this behaviour.
If we denote by $\beta^*(N)$ the position of the maximum of $\chi_\mathcal{O}(\beta)$, also known as the \emph{pseudo-critical value} for the coupling $\beta$, then one expects a scaling relation
\begin{equation}
    \beta^*(N_3) \approx \beta^*(\infty) - b\, N_3^{-\frac{1}{\nu}}
    \label{eq:scaling}
\end{equation}
to hold for large $N_3$.
The critical coupling $\beta^*(\infty)$ at infinite volume, the constant $b$ and the scaling exponent $\nu$ can be obtained by fitting this relation.

The maximum $\chi^{\mathrm{max}}_{\mathcal{O}}(V) = \chi_\mathcal{O}(\beta^*(N_3))$ itself is also expected to scale with a power law,
\begin{equation}
    \chi^{max}_{\mathcal{O}}(V) = c\, N_3^{\frac{\gamma}{\nu}},
\end{equation}
with $c$ a proportionality constant and $\gamma$ the second scaling exponent. 
It is generally argued that at a first-order phase transition one expects $\nu = 1$, while higher-order transitions can give rise to other values.

Another measure of a phase transition is signaled by the change in the distribution of an order parameter while crossing the phase transition line.
In the case of a first-order phase transition, the emergence of the double-peak structure is a strong signal of the co-existence of the phases. The lack of such a distribution and the existence of a distribution different from those which are distant from the transition point could signal the appearance of new critical phenomena. In at least some of the phase transitions (in our case, we observed it for the BP to TT transition), the distribution near the transition changes from Gaussian-like to an asymmetric distribution, with the relative weight of its tails depending sensitively on whether one is above or below the pseudocritical coupling. This observation can be used for an alternative determination of the latter by measuring the skewness of the distribution, which conveniently captures this asymmetry. 
In order to pinpoint the pseudocritical coupling without running simulations at finer and finer spacing in the coupling space, we use \emph{reweighting} to interpolate the $\beta$-dependence of expectation values of observables.
For example, in the case $\beta = \kappa_0$, reweighting takes advantage of the fact that the system at two slightly different values $\kappa_0$ and $\kappa_0'$ will have distributions for the number of vertices $N_0$ that overlap significantly.
Then one can compute an expectation value of $\mathcal{O}$ at coupling $\kappa_0'$ in terms of a reweighted expectation value at coupling $\kappa_0'$,
\begin{equation}
    \langle \mathcal{O} \rangle_{\kappa_0'} = \frac{1}{\mathcal{Z}_{\varepsilon,N}^{\kappa_0'}}\sum_{(\mathcal{T},\mathcal{S},\mathcal{S}^*)} \frac{1}{C_{\mathcal{T}}} \mathcal{O} \,e^{-S_{\varepsilon,N}^{\kappa_0'}} = \frac{\sum_{(\mathcal{T},\mathcal{S},\mathcal{S}^*)} \frac{1}{C_{\mathcal{T}}} \mathcal{O}\, e^{(\kappa_0'-\kappa_0)N_0} \,e^{-S_{\varepsilon,N}^{\kappa_0}}}{\sum_{(\mathcal{T},\mathcal{S},\mathcal{S}^*)} \frac{1}{C_{\mathcal{T}}} e^{(\kappa_0'-\kappa_0)N_0} \,e^{-S_{\varepsilon,N}^{\kappa_0}}} = \frac{\langle\mathcal{O}\, e^{(\kappa_0'-\kappa_0)N_0}\rangle_{\kappa_0}}{\langle e^{(\kappa_0'-\kappa_0)N_0}\rangle_{\kappa_0}}.
\end{equation}
This way with a single simulation at value $\kappa_0$ one obtains estimates for $\langle \mathcal{O} \rangle_{\kappa_0'}$ in a small interval of $\kappa_0'$ around $\kappa_0$.

\section{Phase Diagram of the model}

In this section we report the results of the exploration of the phase diagram for $\beta_{\mathrm{C}} \leq 0$ and $\beta_{\mathrm{L}},\kappa_0 \in \mathbb{R}$.
One of the central questions is whether there are other phases besides the branched polymer (BP) phase and crumpled (CR) phase known from earlier works on DT, whose qualitative properties do not seem to change by the decoration with uniform spanning trees (at $\beta_{\mathrm{L}} = \beta_{\mathrm{C}} = 0$).  
If new phases and phase transitions are present, we expect (at least some of) them to extend to extreme values of the coupling constants, i.e.\ to the boundary of the phase diagram where a combination of $N_0$, $N_{\mathrm{L}}$, $N_{\mathrm{C}}$ is minimized or maximized.
Of particular interest is the corner $\beta_{\mathrm{C}} = \beta_{\mathrm{L}} = -\infty$ corresponding to the triple-tree class, where previous simulations have indicated the presence of a phase transition\footnote{In particular, we wish to highlight the master thesis work \cite{gerstel} of Tom Gerstel and his Monte Carlo implementation available at \url{https://github.com/TomGerstel/dynamical-triangulations-3d}.}.
Moreover, as explained in Section~\ref{sec:model}, at the extreme values $\kappa_0 = \pm\infty$ the triangulations with maximal number (melonic) or minimal number (crystallizations) of vertices are also of interest, because they are within closer combinatorial reach than general 3-sphere triangulations.
We should, however, acknowledge that the ergodicity of our Markov chain is only granted at finite values of the coupling constants.
Because of this, we focus on values on the boundary of the finite but large box $(\kappa_0,\beta_{\mathrm{L}},\beta_{\mathrm{C}}) \in [-10,10]^3$ in the phase diagram.
As we will briefly discuss in the outlook in Section~\ref{sec:outlook}, the top half of this box corresponding to $\beta_{\mathrm{C}} > 0$ has a rich and rather complicated structure, so to limit our scope we focus on the lower half $\beta_{\mathrm{C}} \leq 0$.

\subsection{Observables}
Finding suitable order parameters (OP) is generally difficult in random geometry models, but luckily several observables come naturally with the model at hand. One of the natural OPs in standard DT is the number of vertices $N_0$ in the triangulation. In this context, DT is known to exhibit two distinct phases: the crumpled phase (CR), in which triangulations have relatively few vertices while some vertices and edges with very high coordination number are present; and the branched-polymer phase (BP), which features triangulations with many vertices and the tetrahedra are organized in a tree-like fashion. 
The maximal vertex coordination number $c_{\mathrm{v}}^{\mathrm{max}}$ and maximal edge coordination number $c_{\mathrm{e}}^{\mathrm{max}}$, counting the maximal number of tetrahedra incident to a single vertex or edge, make for useful OPs for this transition. The introduction of the two new couplings in our model naturally expands the range of observables. Since the number of components ($N_{\mathrm{C}}$) and the number of loops ($N_{\mathrm{L}}$) in the middle graph are dual to the new couplings, they are expected to show strong dependence on the couplings and thus to provide insight into potential phase transitions induced by them. 

\subsection{General Structure at $\beta_{\mathrm{C}} \leq 0$}

\begin{figure}[ht]
    \centering
    \includegraphics[width = .7\textwidth]{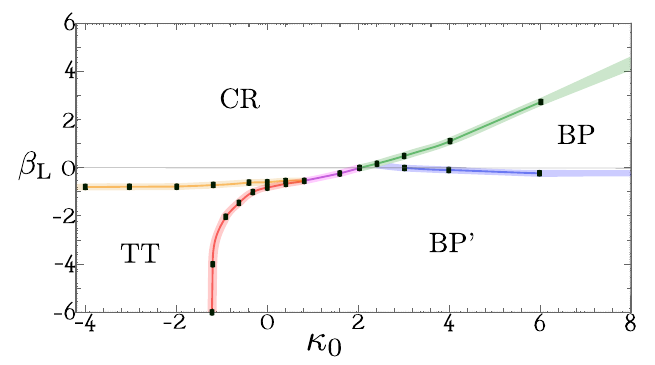}
    \caption{A slice of the phase diagram at $\beta_{\mathrm{C}} = 0$ for $N_3 = 4k$. The dots indicate measurement points; their size reflects the uncertainty of the position.}
    \label{fig:PD_beta0}
\end{figure}

To gain a rough understanding of the structure of the new phase diagram and prepare the ground for a more detailed scaling analysis, we performed an initial set of simulations at fixed volume $N_3 = 4000$ (denoted $4\mathrm{k}$ in the following). This first mapping is restricted to the plane $\beta_{\mathrm{C}} = 0$, varying $\kappa_0$ and $\beta_{\mathrm{L}}$. 

This slice includes the case $\beta_{\mathrm{L}} = 0$, corresponding to standard DT decorated with uniform spanning trees, which is known to exhibit a transition between the branched polymer (BP) and crumpled (CR) phases. By locating peaks in the susceptibilities of various observables, we are led to the tentative phase diagram shown in Fig.~\ref{fig:PD_beta0}. It reveals a rich structure with four potential phases and five possible pseudo-critical transition lines.

In the region of positive $\kappa_0$ and negative $\beta_{\mathrm{L}}$, we identify a regime labeled BP', which shares many characteristics with the standard BP phase but has a notably lower loop number $N_{\mathrm{L}}$. In the bottom-left corner, at negative $\kappa_0$ and $\beta_{\mathrm{L}}$, a seemingly new phase appears, tentatively referred to as the triple-tree (TT) phase. It exhibits qualitative features distinct from both the BP and CR phases. 

Most of the transitions appear at small positive or negative values of the couplings. While the present section focuses on $\beta_{\mathrm{C}} \leq 0$, it is worth noting that the structure depicted in Fig.~\ref{fig:PD_beta0} extends slightly into the region $\beta_{\mathrm{C}} > 0$. New structures appear around $\beta_{\mathrm{C}} \approx 2$, particularly above the TT and BP phases, while the CR phase seems to extend to larger $\beta_{\mathrm{L}}$ values. A more detailed analysis will be presented elsewhere; for further remarks, see Section~\ref{sec:outlook}.

\begin{figure}[h]
    \centering
    \includegraphics[width=0.3\textwidth]{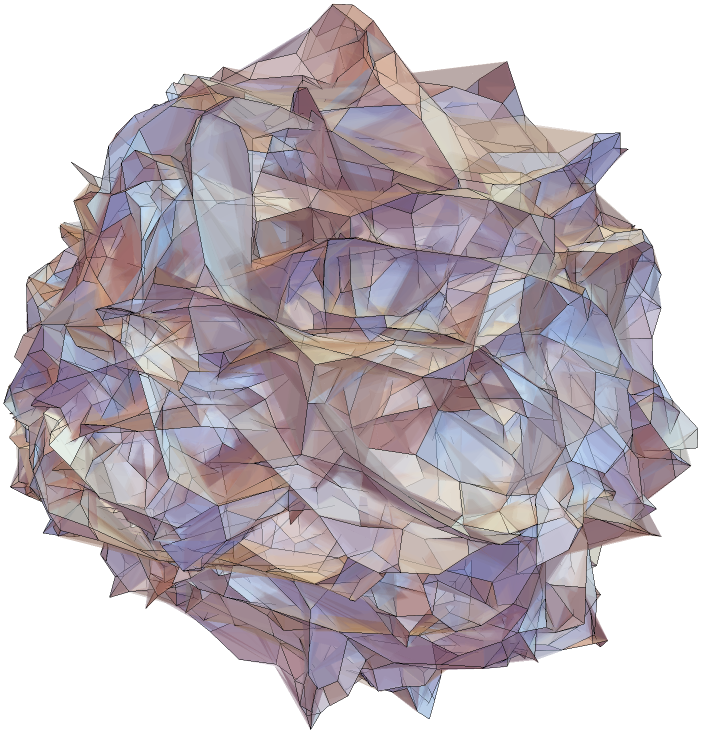}
    \includegraphics[width=0.6\textwidth]{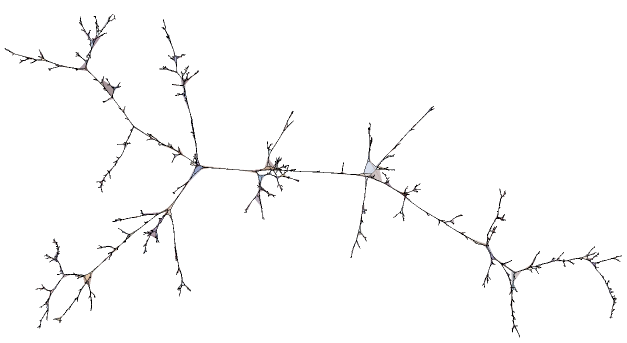}
    \caption{Comparison of the dual graphs of triangulations with 8000 tetrahedra in the crumpled (CR) and branched polymer (BP) phases at coordinates $(\kappa_0,\beta_L,\beta_C) = (-10,10,-10)$ and $(10,-10,-10)$ respectively.}
    \label{fig:dual_g_OLD}
\end{figure}

One way to appreciate the differences between the phases is by visualizing the geometry of randomly sampled triangulations from the Boltzmann ensemble. Fig.~\ref{fig:dual_g_OLD} shows spring-electrical embeddings in three dimensions of the dual graphs of typical triangulations in the already known CR and BP phases. Since they represent the dual graph, lines in the figure correspond to faces of the triangulation, while the colored faces dual to a subset of edges in the triangulation are included for better visibility of the structure.

\begin{figure}[h]
    \centering
        \includegraphics[width=0.5\textwidth]{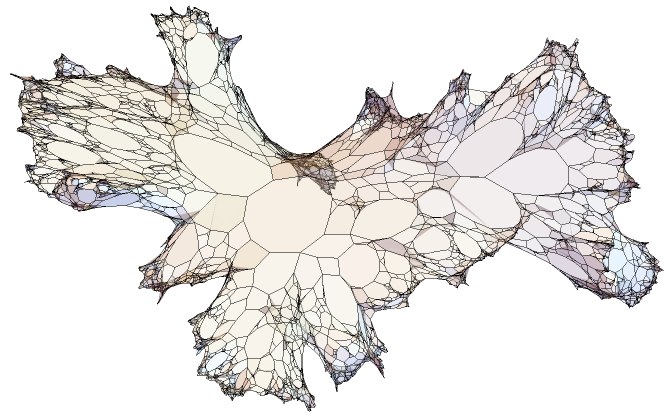}
    \caption{Dual graph of a typical configuration with 8000 tetrahedra at the triple-tree (TT) phase at coordinate $(\kappa_0,\beta_L,\beta_C) = (-10,-10,-10)$.}
    \label{fig:dual_g_TT}
\end{figure}

The embeddings shown in Fig.~\ref{fig:dual_g_OLD} and ~\ref{fig:dual_g_TT} correspond to triangulations at extreme values of the couplings within the $\beta_{\mathrm{C}} = -10$ slice: $(\kappa_0, \beta_{\mathrm{L}}) = (-10, 10)$ for CR, $(-10, -10)$ for TT, and $(10, -10)$ for BP. The difference in the characteristics of the graphs is striking. The BP phase (both BP and BP') features a tree-like graph, while the embedding of the graph of the CR phase shows a strongly connected crumpled structure. On the other hand, the new TT phase (shown in Fig.~\ref{fig:dual_g_TT}) features a structure that resembles that of a planar graph. 

\begin{figure}[h]
    \centering
    \includegraphics[width = 0.45\textwidth]{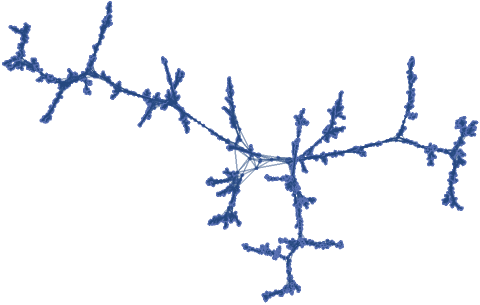}
    \caption{The vertex graph in the branched polymer phase of a typical triangulation with 8000 tetrahedra.}
    \label{fig:vert_g}
\end{figure}

While some local observables (e.g., $N_0$, $N_{\mathrm{L}}$, $N_{\mathrm{C}}$) vary within a phase, the main qualitative features of both the dual and vertex graphs remain consistent throughout a given phase. For example, the vertex graph becomes  fully connected in the CR and TT phases only when the number of vertices is small, which happens at extreme $\kappa_0$ values.
The vertex graph in the BP phase, shown in Fig.~\ref{fig:vert_g}, mirrors the tree-like structure of the dual graph. 

\begin{figure}[h!]
    \centering
    \includegraphics[width = 0.7\textwidth]{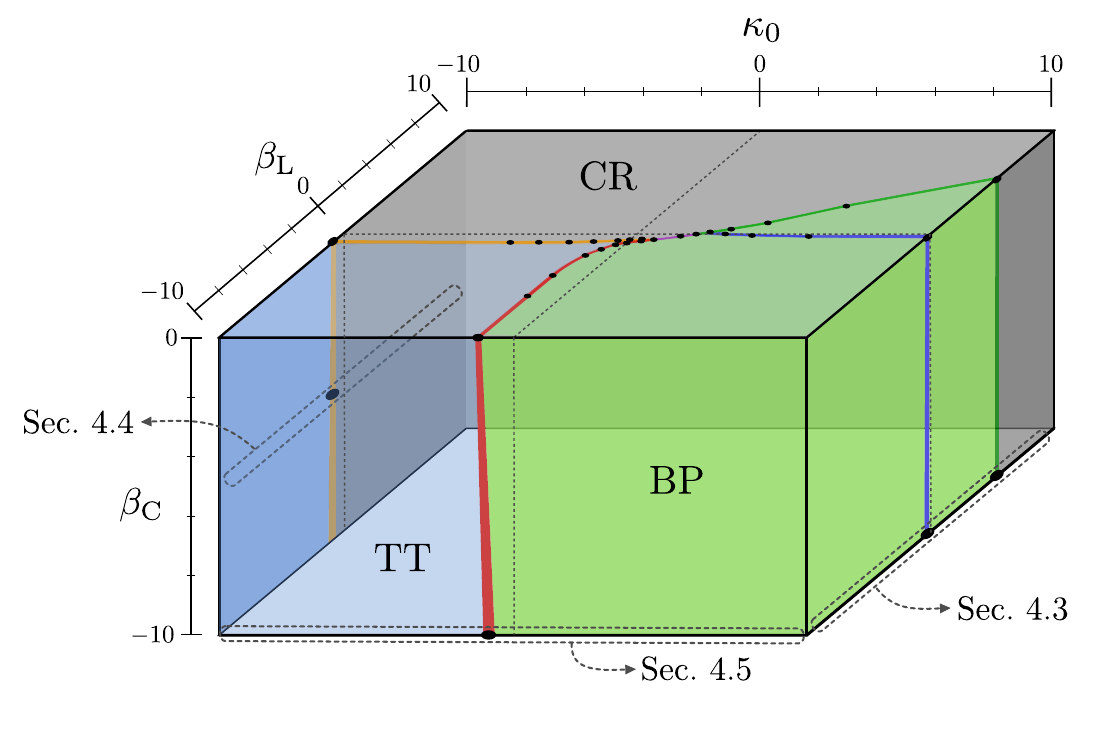}
    \caption{Schematic illustration of the three-dimensional phase diagram restricted to the region $\beta_{\mathrm{C}} \leq 0$. The three phases are tentatively identified as crumpled (CR, gray), branched polymer (BP, green), and triple-tree (TT, blue).}
    \label{fig:3d_half}
\end{figure}

In the next three subsections we present a quantitative analysis of the potential phase transitions, focusing on lines on the boundary of the box at fixed negative $\beta_{\mathrm{C}}$. 
A schematic of the inferred three-dimensional phase structure is shown in Fig.~\ref{fig:3d_half}, which also indicates the lines under consideration.

\subsection{The BP-CR Transition: Results at $\kappa_0 = 10$, $\beta_{\mathrm{C}} = -10$}\label{sec:BPCR}

In this subsection we investigate the BP-CR transition at fixed $\kappa_0 = 10$ and $\beta_{\mathrm{C}} = -10$, corresponding to one of the edges of the parameter space (bottom-right of Fig.~\ref{fig:3d_half}). The loop coupling $\beta_{\mathrm{L}}$ is varied in the range $\beta_{\mathrm{L}} \in [-10,10]$, with simulations performed for volumes $N_3$ ranging from $40$ up to $6000$. 

\begin{figure}[t]
    \centering
    \includegraphics[width = 0.7\textwidth]{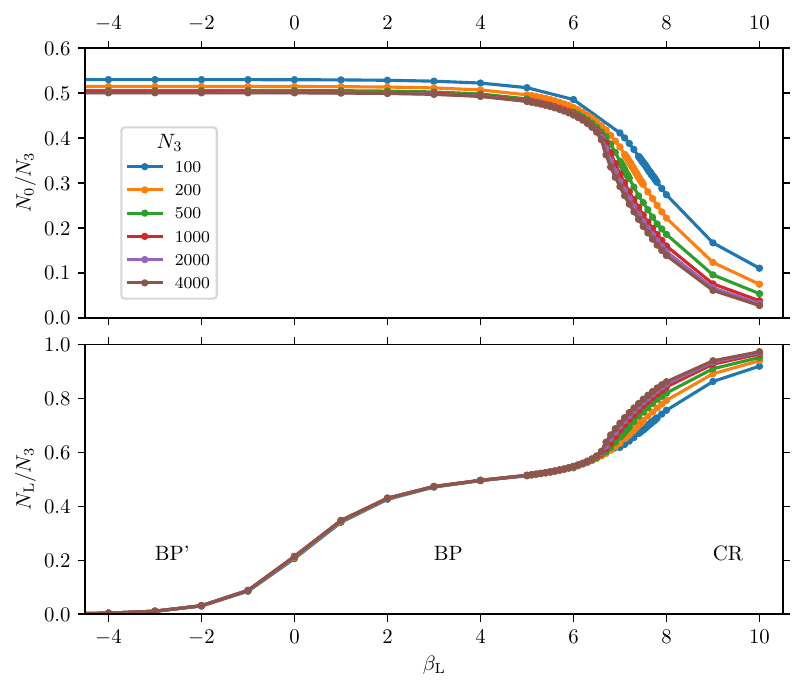}
    \caption{The volume-normalized number of vertices $N_0$ (left) and the number of loops $N_{\mathrm{L}}$ (right) as a function of $\beta_{\mathrm{L}}$. The right panels shown an enlarged view of the vicinity of the BP-CR transition.}
    \label{fig:ops_k10}
\end{figure}

Fig.~\ref{fig:ops_k10} shows the volume-normalized mean number of vertices $N_0$ and the number of loops $N_{\mathrm{L}}$ as a function of $\beta_{\mathrm{L}}$.
A sharp transition at $\beta_{\mathrm{L}} \approx 7$ is visible in $N_0$, signaling the BP-CR transition: $N_0$ drops rapidly from its maximal value $\approx N_3/2$ to a minimal value indicative of collapsed geometries.
Correspondingly, $N_{\mathrm{L}}$ displays a strong dependence on $\beta_{\mathrm{L}}$ at this point, transitioning from $N_{\mathrm{L}}\approx N_3/2$ to $N_{\mathrm{L}} \approx N_3$.
Additionally, $N_{\mathrm{L}}$ shows a secondary jump near $\beta_{\mathrm{L}} \approx 0$ deep within the BP phase, while $N_0$ remains largely unaffected.
This behaviour may stem from lattice artifacts or changes in substructure, supported by the emergence of two distinct plateaus for the order parameter $N_{\mathrm{L}}$ within the BP regime.

\begin{figure}[h]
    \centering
    \includegraphics[width = 0.8\textwidth]{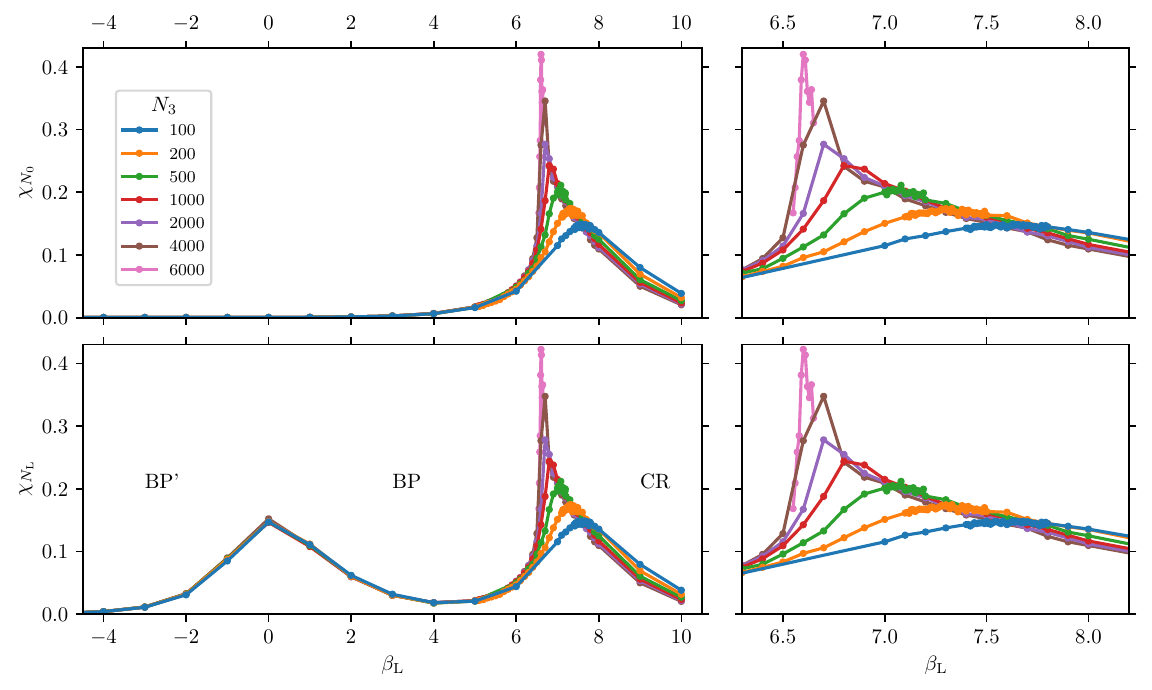}
    \caption{The susceptibility of $N_0$ (top) and $N_{L}$ (bottom) as a function of $\beta_{\mathrm{L}}$. The figures on the right show magnifications of the vicinity of the BP-CR transition. Note that in this regime $N_0 \approx N_{\mathrm{L}}$, explaining the close resemblance of the two plots.}
    \label{fig:vars_k10}
\end{figure}

The susceptibilities $\chi_{N_0}$ and $\chi_{N_{\mathrm{L}}}$ in Fig.~\ref{fig:vars_k10} exhibit clear peaks near $\beta_{\mathrm{L}} \approx 7$, consistent with a phase transition. 
The additional peak in $\chi_{N_{\mathrm{L}}}$ near $\beta_{\mathrm{L}} \approx 0$ does not scale with volume, suggesting that it is not a true phase transition but a structural crossover.
A closer examination of the structure of the geometries revealed that the regions (BP' and BP) on both sides of the crossover have all the characteristics of branched polymers, corroborated by a preliminary study of their Hausdorff and spectral dimensions. Apart from a shift in the number of loops, none of our measurements provide evidence for a change in the macroscopic geometry between these two regions.

\begin{figure}[t]
    \centering
    \includegraphics[width = 0.8\textwidth]{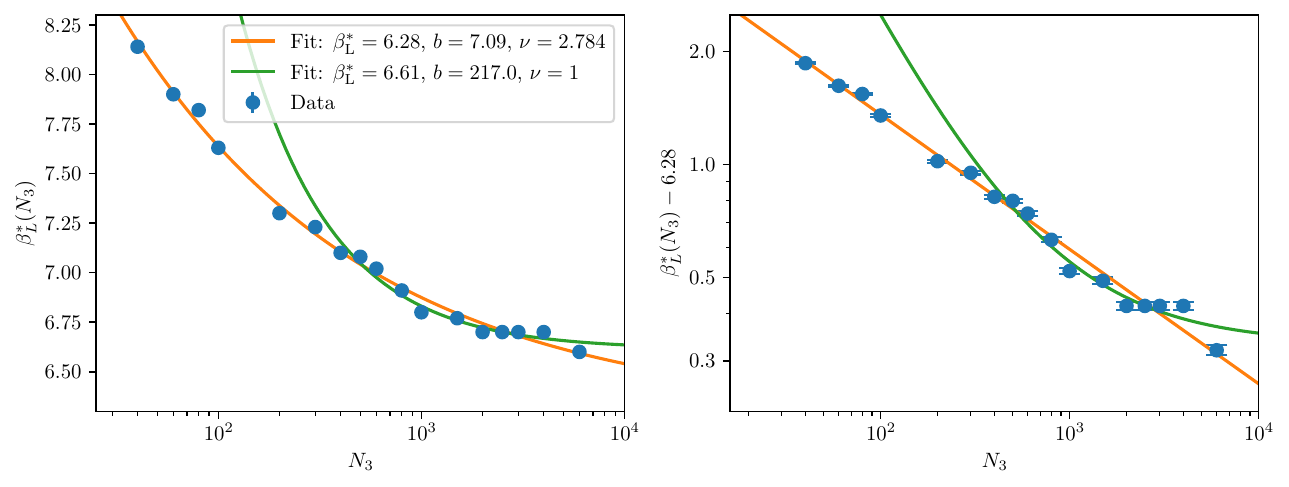}
    \caption{The pseudo-critical values $\beta^*_{\mathrm{L}}(N_3)$ of the loop coupling calculated from the positions of the peaks of $\chi_{N_0}$. Fits to the scaling relation $\beta^*_{\mathrm{L}}(N_3) = \beta^*_{\mathrm{L}} - b\, N_3^{-1/\nu}$ are shown for $\nu = 1$ (green) as well as unrestricted $\nu$ (orange).}
    \label{fig:beta_L_v_AB}
\end{figure}
\begin{figure}[h!]
    \centering
    \includegraphics[width = .9\textwidth]{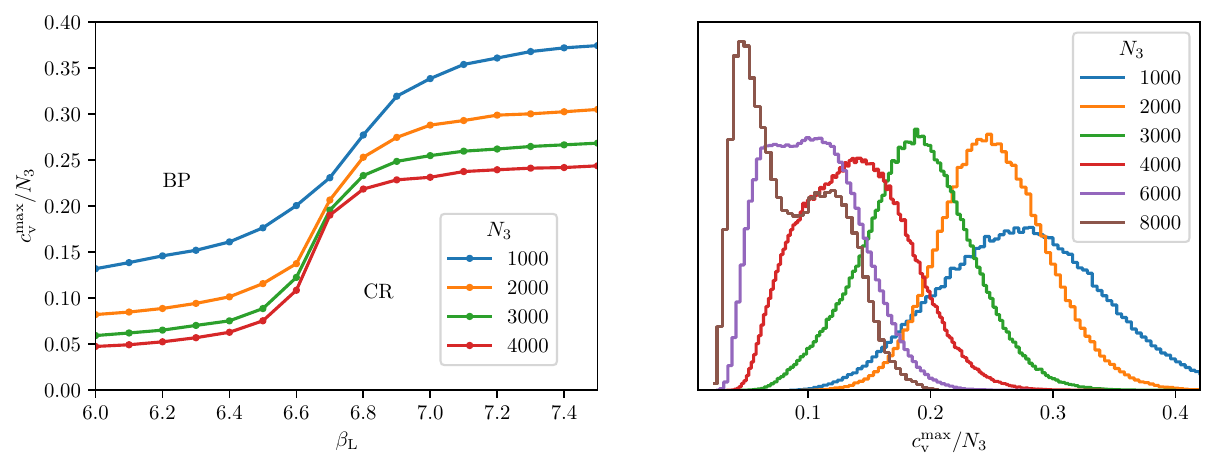}
    \includegraphics[width = 0.7\textwidth]{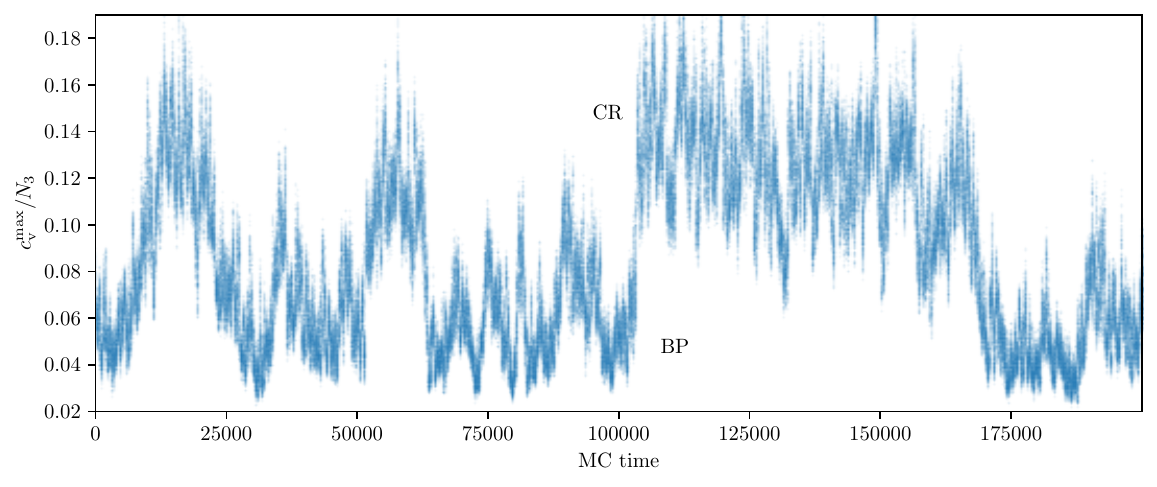}
    \caption{The mean maximal vertex coordination number $c_{\mathrm{v}}^{\mathrm{max}}$ as a function of $\beta_{\mathrm{L}}$ (top-left) and its histograms close to the critical point for a few volumes (top-right). A double peak structure can be seen to appear for the largest volumes. The transitions are also visible in the Monte Carlo trace for $N_3 = 8000$ (bottom).} 
    \label{fig:double_peak}
\end{figure}

\begin{figure}[h!]
    \centering
    \includegraphics[width = 1.03\textwidth]{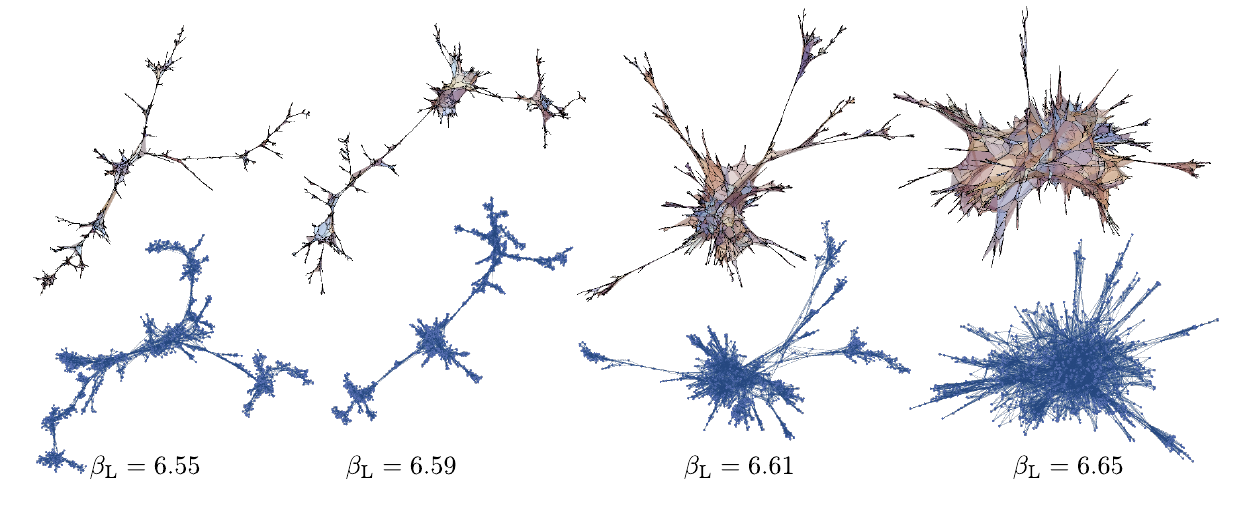}
    \caption{The dual graphs (top) and vertex graphs (bottom) of typical geometries across the BP-CR transition for $N_3 = 6\mathrm{k}$.}
    \label{fig:simps_b_a}
\end{figure}

A finite-size scaling analysis of the peak in $\chi_{N_0}$ was conducted to extract the critical coupling $\beta_{\mathrm{L}}^{*}$ associated with the BP-CR transition. 
Fig.~\ref{fig:beta_L_v_AB} shows the positions of the peaks as function of the system size, while the fits were performed on the basis of the scaling relation of equation \eqref{eq:scaling}. The green curve corresponds to a fit with a fixed critical exponent $\nu = 1$, while for the orange curve this exponent is also fitted, leading to $\nu = 2.8 \pm 0.2$. The corresponding values for the fitted critical coupling are $\beta_{\mathrm{L}}^* = 6.3 \pm 0.1$ and $\beta_{\mathrm{L}}^* = 6.6 \pm 0.1$, respectively. 

Although the scaling analysis hints at a critical exponent $\nu \neq 1$, the fit quality depends sensitively on the range of system sizes included, which means that the finite-size effects are not well under control.
A more robust signal of a first-order nature of the transition is the appearance close to the transition of a double-peak structure in the histogram of an order parameter. 
For this we examine the maximal vertex coordination number $c_{\mathrm{v}}^{\mathrm{max}}$ that shows a strong dependence on $\beta_{\mathrm{L}}$, see Fig.~\ref{fig:double_peak}.
The top-right panel shows the appearance of the double-peak structure for the largest volumes of our dataset, while the bottom panel shows the corresponding Monte Carlo trace of $c_{\mathrm{v}}^{\mathrm{max}}$. Multiple jumps between the two phases signal the coexistence of the two phases at the transition. The signal, however, is considerably weaker than in the case of BP-CR transition in regular DT. The vertex graphs and dual graphs of typical random geometries are shown in Fig.~\ref{fig:simps_b_a}.

In conclusion, we observe a clear phase transition at $\kappa_0 = 10$, which is the regular DT phase transition, known to be first-order. The phase transition appears weaker than in the case of vanilla DT, but the distribution of the maximal vertex order signals the double-peak structure. The phase transition at this location can be safely characterized as first-order. However, the addition of a new coupling parameter obscures this behavior, as the emergence of the double peak is not as strong as in vanilla DT. We cannot exclude the potential weakening of the phase transition at infinite $\kappa_0$. Additionally, a sign of a phase transition was observed without any scaling of our current observables inside the Branched Polymer phase, suggesting a sub-phase of BP with a slightly different microscopic setup.

\subsection{The TT-CR Transition: Results at $\kappa_0 = -10$, $\beta_{\mathrm{C}} = -5$}

\begin{figure}[h!]
    \centering
    \includegraphics[width = 0.6\textwidth]{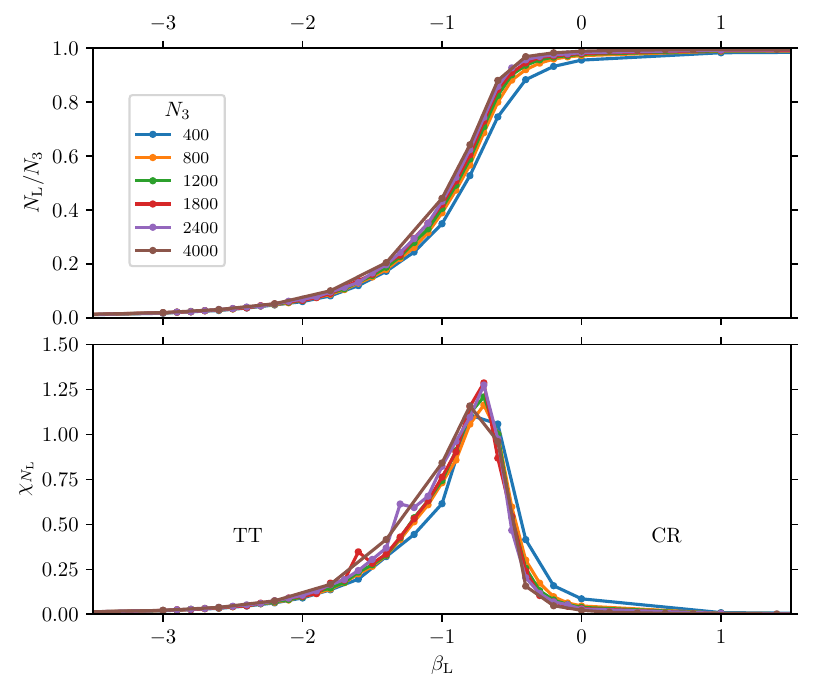}
    \caption{Volume-normalized mean value of $N_{L}$ (top) and the corresponding susceptibility (below).}
    \label{fig:k0_m10_nl}
\end{figure}

The other side of the phase diagram in Fig.~\ref{fig:3d_half}, corresponding to the fixed values $\kappa_0 = -10$ and $\beta_{\mathrm{C}} = -5$, presents another potential phase transition. 
The geometry here is highly asymmetric with respect to the sizes of the spanning trees. In this region of the phase diagram, typical geometries exhibit a very low number of vertices, which drastically reduces the degrees of freedom in the vertex trees of the triangulation. In the extreme case of $\kappa_0 = -\infty$, the number of vertices approaches $N_0 = 4$, corresponding to triangulations known as crystallizations. As $\beta_{\mathrm{L}}$ is varied, the middle graph transitions from geometries with few loops to those with a large number of loops, as supported by the data shown in Fig.~\ref{fig:k0_m10_nl}.

\begin{figure}[h!]
    \centering
    \includegraphics[width = 0.6\textwidth]{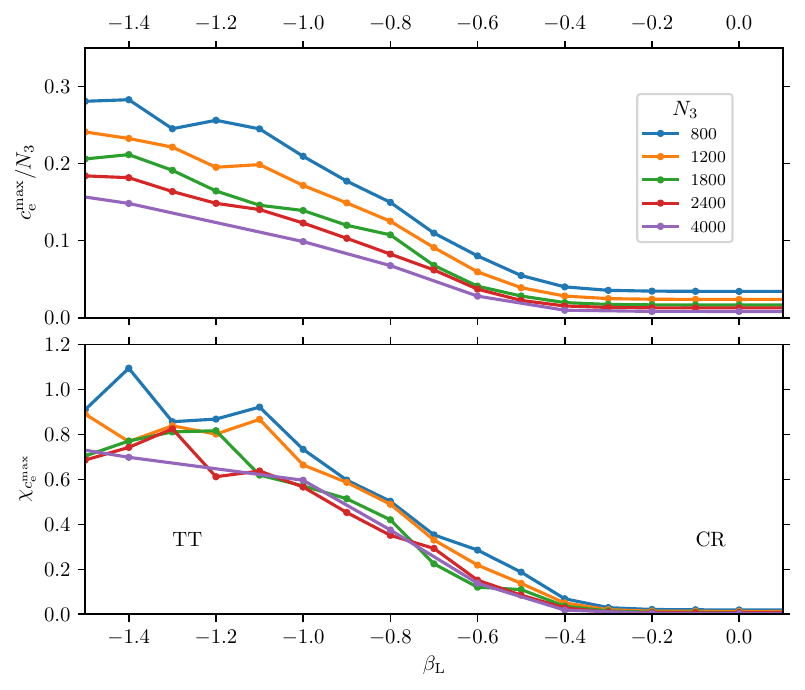}
    \caption{The mean maximal edge coordination number $c_{\mathrm{e}}^{\mathrm{max}}$ normalized by volume (top) and the corresponding susceptibility (below).}
    \label{fig:k0_m10_cemax}
\end{figure}

The top panel of Fig.~\ref{fig:k0_m10_nl} reveals a sharp transition: for large negative values of $\beta_{\mathrm{L}}$, the middle graph contains a minimal number of loops, while for large positive values, the number of loops grows to approach the total volume. However, although the susceptibility shows a peak at around $\beta_{\mathrm{L}}=-0.8$, no scaling with increasing system size is visible. 

\begin{figure}[h!]
    \centering
    \includegraphics[width=\linewidth]{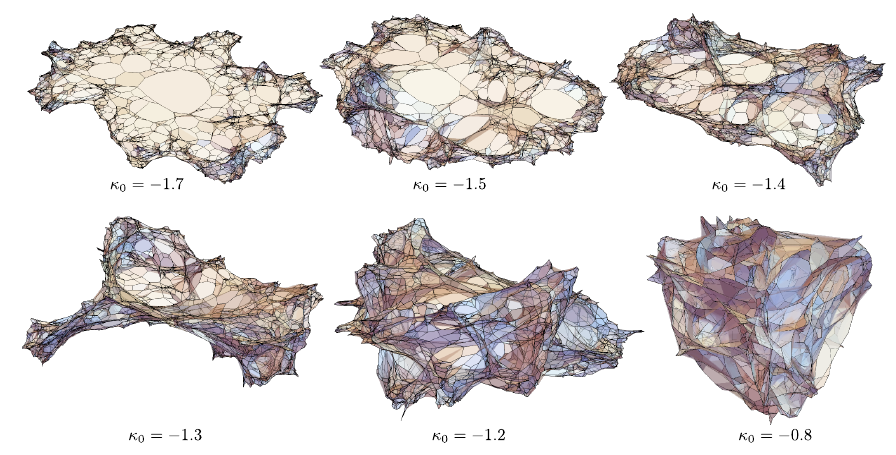}
    \caption{Dual graphs of typical geometries across the TT-CR transition for $N_3 = 6k$.}
    \label{fig:simps_tt_6k}
\end{figure}

Another observable, the maximal edge coordination number $c_{\mathrm{e}}^{\mathrm{max}}$, shows similarly strong dependence on $\beta_{\mathrm{L}}$ but also here no clear scaling of the susceptibility is seen (see Fig.~\ref{fig:k0_m10_cemax}).
The presence of macroscopic coordination numbers ($\approx 0.2 N_3$) in the TT phase to small coordination numbers in the CR phase, suggests that the geometries in the two phases are drastically different. 
This is in contrast with the (non-)transition at $\beta_{\mathrm{L}} \approx 0$ observed in the BP phase in the previous subsection, where no qualitative change in the geometry was observed.
In the TT-CR case the differences are already evident visually, see Fig.~\ref{fig:simps_tt_6k}.
The geometries observed across the phase transition, evolve from crumpled to planar-looking structures. No tree-like features were observed in either the vertex or the dual graph of the triangulations. Also preliminary measurements of Hausdorff and spectral dimensions of the geometries show qualitative differences. A detailed analysis of the TT phase and the crystallization limit  of the phase diagram will be part of an upcoming study, but there is already good reason to believe that there is a phase transition between the TT and BP phases (see next subsection), thereby rendering the TT phase a distinct phase from those known in Dynamical Triangulations.

 \subsection{The TT-BP Transition: Results at $\beta_{\mathrm{L}} = -10$, $\beta_{\mathrm{C}} = -10$}

The front side of Fig. \ref{fig:3d_half} features one of the most interesting phase transitions in the enlarged coupling space, and was the primary focus of this study. Simulations with couplings $\beta_{\mathrm{C}} = \beta_{\mathrm{L}} = -10$ force the middle graph to have only very few connected components and loops, restricting its structure to be close to a tree. Our algorithm is not necessarily the most efficient way to sample geometries featuring exact triple-trees ($N_{\mathrm{L}}=0$, $N_{\mathrm{C}}=1$), since the dynamical nature of the Markov chain allows for deviations. However, by setting the couplings to large negative values, we can effectively and dynamically enforce the desired structures in the middle graph.

\begin{figure}[t] \centering 
\includegraphics[width = 0.9\textwidth]{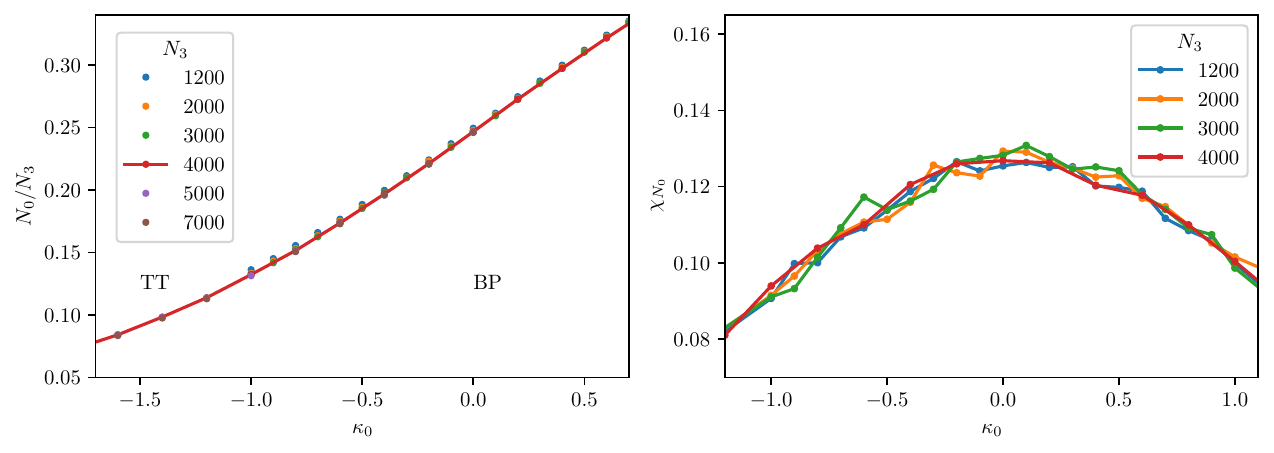} 
\includegraphics[width = 0.75\textwidth]{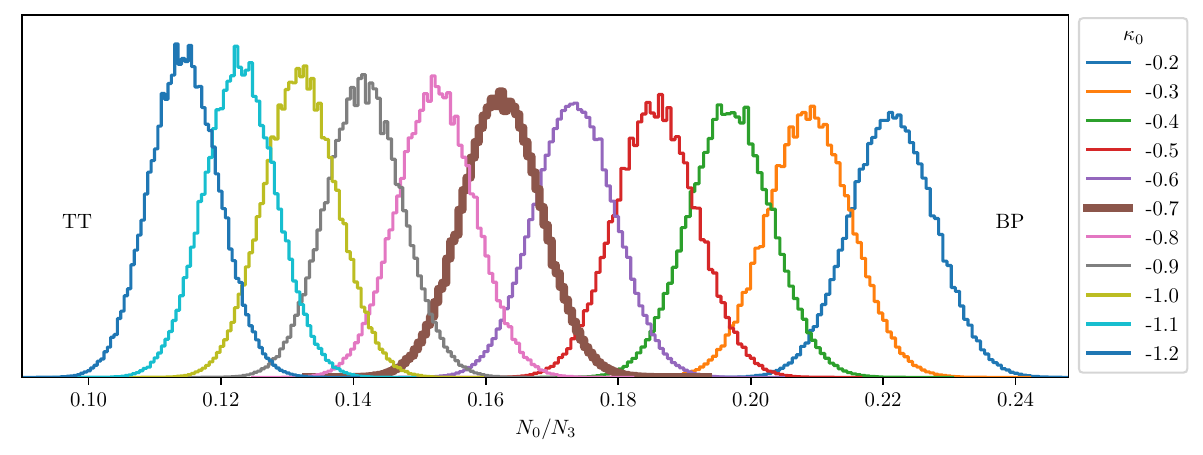} \caption{
The mean number of vertices $N_0/N_3$ (top-left), its susceptibility $\chi_{N_0}$ (top-right), and histograms of $N_0/N_3$ at $N_3 = 3\mathrm{k}$ for different values of $\kappa_0$ (bottom).} \label{fig:n0_hist_tt} \end{figure}

In comparison with the BP phase, the TT phase has fewer vertices than the BP phase, see the top-left plot of Fig.~\ref{fig:n0_hist_tt}, but no jump is visible.
The presence of vertices and edges with large coordination numbers is more distinctive for the TT phase, yielding a geometric structure that is less elongated than the branched polymers but also less singular than the crumpled phase. 
The branched polymer's elongated structure cannot support vertices and edges with very large coordination number, thus we expect a jump in these quantities at a phase transition.
Judging by the top-left plot of Fig.~\ref{fig:reweight} below, this is indeed happens but the transition is not nearly as sharp as in the BP-CR case of Section~\ref{sec:BPCR}.

To start our analysis we examine histograms at a fixed volume $N_3 = 3\mathrm{k}$. Extensive observables, like the counts of local substructures, are expected to have an approximately normal distribution deep inside the phases, as is the case for the number of vertices $N_0$ shown in Fig.~\ref{fig:n0_hist_tt}. 
This behaviour appears to persist across the tentative transition.
Since the susceptibility $\chi_{N_0}$ does not present noticeable scaling behaviour either, this suggests $N_0$ is unsuitable as an order parameter for the transition. 

\begin{figure}[h!] 
\centering 
\includegraphics[width = 0.95\textwidth]{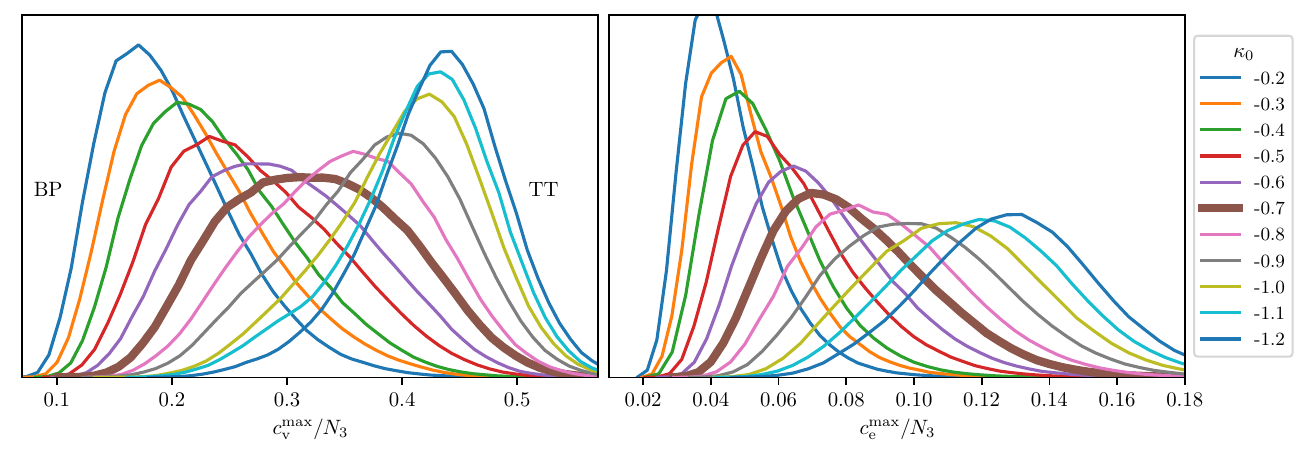} \caption{Histograms of the maximal vertex (left) and edge (right) coordination numbers for different values of $\kappa_0$ for fixed $N_3 = 3000$.} \label{fig:hist_tt} \end{figure}

\begin{figure}[h!] \centering \includegraphics[width = 0.5\textwidth]{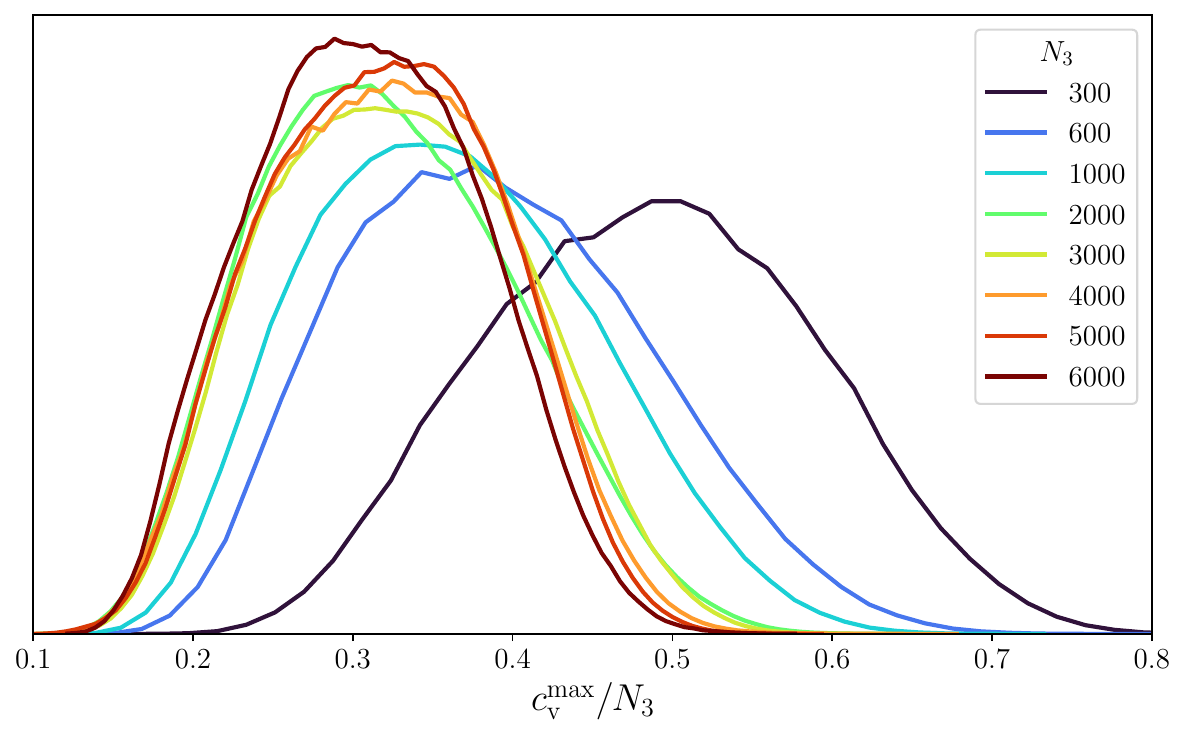}
\includegraphics[width = 0.7\textwidth]{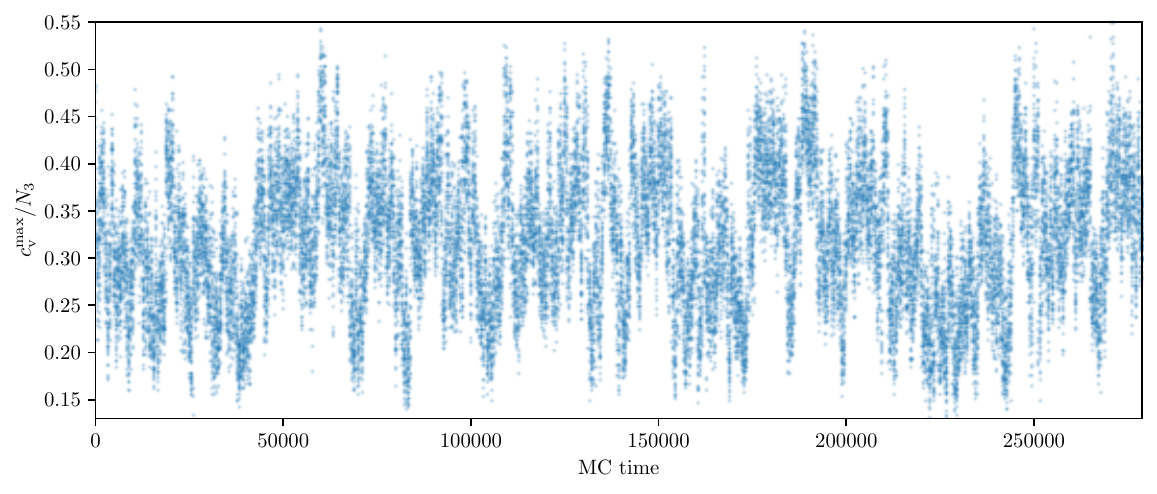} 
\caption{Histograms (top) of the maximal vertex coordination number for $\kappa_0$ close to $\kappa_0^{*}(N_3)$, together with a Monte Carlo trace (bottom) for $N_3 = 6k$.} \label{fig:6hist} \end{figure}

In contrast, the histograms of the maximal vertex and edge coordination numbers $c_{\mathrm{v}}^{\mathrm{max}}$ and $c_{\mathrm{e}}^{\mathrm{max}}$ exhibit qualitative differences compared to $N_0$, see Fig.~\ref{fig:hist_tt} for $N_3=3\mathrm{k}$.
Away from the transition the distribution in the BP phase is relatively long-tailed on the right and in the TT phase it is long-tailed on the left, while close to the transition an approximately symmetric distribution is found.
This allows us to select a pseudo-critical coupling $\kappa^{*}_0(N_3)$ where $c_{\mathrm{v}}^{\mathrm{max}}$ is closest to being symmetric.
This was done for a range of different volumes, and the histograms of $c_{\mathrm{v}}^{\mathrm{max}}$ for $\kappa_0 \approx \kappa^{*}_0(N_3)$ are shown in Fig.~\ref{fig:6hist}.
There is no signal of a double-peak structure.
Also the Monte Carlo history of $c_{\mathrm{v}}^{\mathrm{max}}$ for one of the largest volumes $N_3= 6\mathrm{k}$ shows no strong jumps between metastable states, as it did in the case of the BP-CR transition.

\begin{figure}[h!] \centering 
\includegraphics[width=\linewidth]{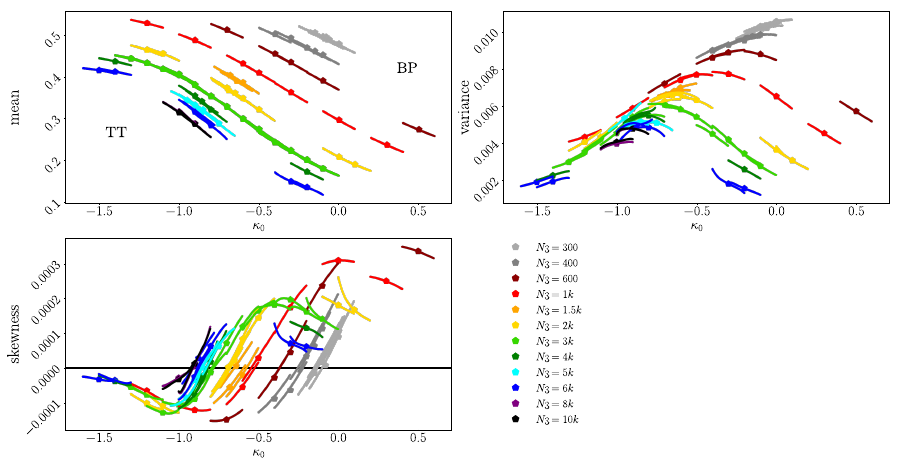}
\caption{Reweighted mean (top-left), variance (top-right) and skewness (bottom-left) of the maximal vertex coordination number $c_{\mathrm{v}}^{\mathrm{max}} / N_3$ for various volumes.} \label{fig:reweight} \end{figure}

With this is mind we proceed with a quantitative analysis of scaling near the phase transition. Fig. \ref{fig:reweight} presents the first three moments of the volume-normalized maximal vertex coordination number $c_{\mathrm{v}}^{\mathrm{max}} / N_3$. The mean values in the top-left panel clearly indicate a scaling in the position of the phase transition, as increasing the volume results in the values shifting less and less.
The variance in the top-right panel shows a peak that does become sharper with increasing volume but not very rapidly.
This means that the height $\chi^*_{c^{\mathrm{max}}_{\mathrm{v}}}$ of the peak can be accurately determined, but its position has significant uncertainty. 
A more robust determination of the pseudo-critical coupling $\kappa_0^*(N_3)$ is obtained by examining the skewness in the bottom-left panel and estimating the location where it crosses the zero line.
Thanks to reweighting this location can be determined rather accurately.

\begin{figure}[h] \centering 
\includegraphics[width = \linewidth]{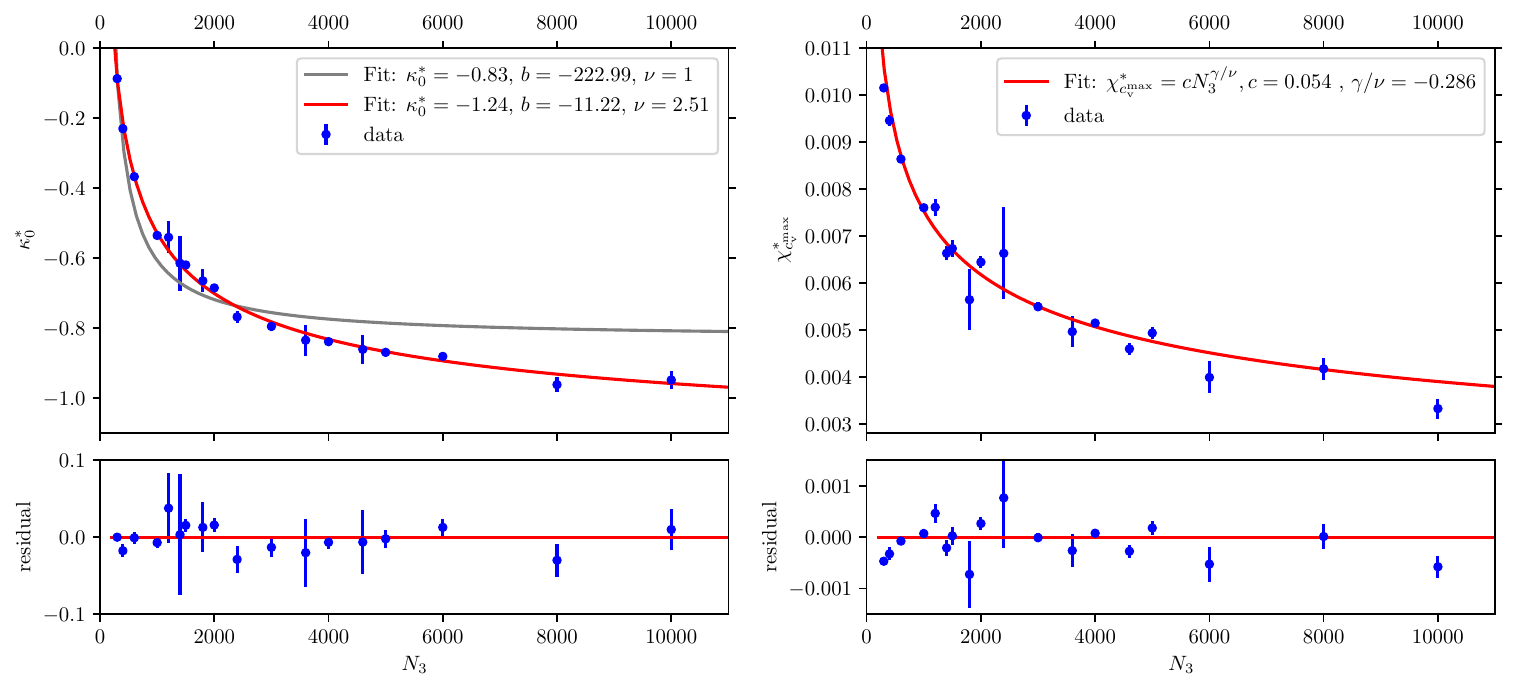}
\caption{The left plot shows the pseudocritical coupling $\kappa_0^*(N_3)$ as extracted from the axis intersection of the skewness of $c_{\mathrm{v}}^{\mathrm{max}}/N_3$ in Fig.~\ref{fig:reweight} together with a best fits of the ansatz \eqref{eq:nu_scaling} for $\nu = 1$ fixed (gray) and freely fitted exponent $\nu$ (red). The right plot shows $\chi^*_{c_{\mathrm{v}}^{\mathrm{max}}}(N_3)$ extracted from the maximum of the variance in Fig.~\ref{fig:reweight} together with the best fit of \eqref{eq:gammascaling}.} \label{fig:k0_tt} \end{figure}

We then perform a finite-size scaling fit as in \eqref{eq:scaling},
\begin{equation}
\kappa_0^{*}(N_3) \approx \kappa_0^*(\infty) - b \cdot N_3^{-\frac{1}{\nu}}, \label{eq:nu_scaling} 
\end{equation}
which is shown in the left plot in Fig.~\ref{fig:k0_tt}.
The parameters were determined to be $\kappa_0^{*}(\infty) = -1.244 \pm 0.036$ and $\nu = 2.51 \pm 0.12$. It is evident from the figure that $\nu = 1$ does not give a plausible fit and this fit does does not significantly improve when considering only larger volumes, so we can conclude $\nu \neq 1$ with good confidence.

The maximum susceptibility $\chi^*_{c_{\mathrm{v}}^{\mathrm{max}}}$ is fitted to the scaling relation
\begin{equation} \chi^*_{c_{\mathrm{v}}^{\mathrm{max}}}(N_3) \approx c\,  N_3^{\gamma/\nu}, \label{eq:gammascaling}\end{equation}
where the scaling exponent $\gamma/\nu$ is normalized with respect to exponent $\nu$ from \eqref{eq:nu_scaling}. 
As shown in the right plot of Fig.~\ref{fig:k0_tt}, the data is consistent with a scaling relation with exponent $\frac{\gamma}{\nu} = -0.286 \pm 0.006$. Although not conclusive, these observations strengthen the claim that the TT-BP transition is continuous.

\begin{figure}[h] \centering 
\includegraphics[width=.9\linewidth]{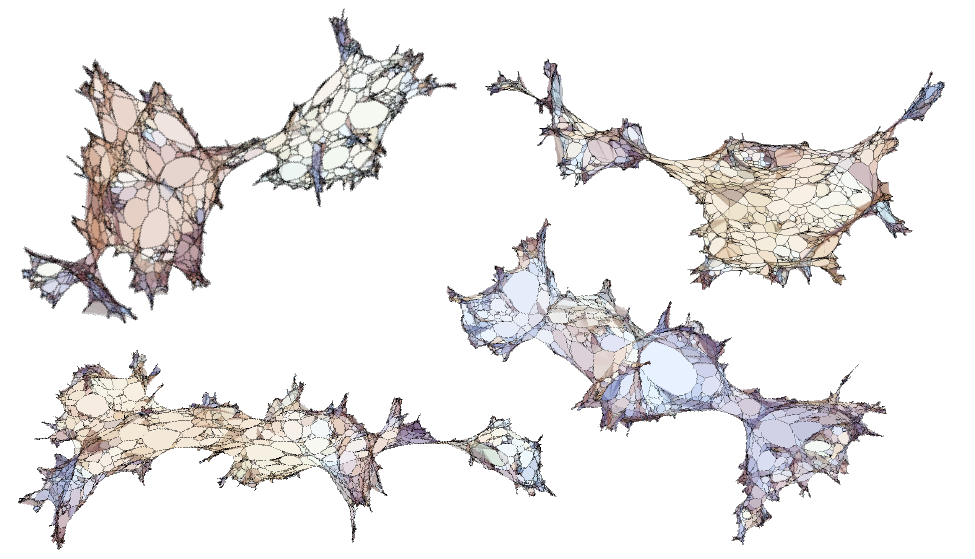}
\caption{Dual graphs of typical random geometries at the pseudocritical point ($\kappa_0 = -0.877$) of the TT-BP transition for $N_3 = 6000$.} \label{fig:config_6k_crit} \end{figure}

This leaves open the possibility of a new universality class of random geometry at the critical point. 
To appreciate this possibility, we present in Fig.~\ref{fig:config_6k_crit} several snapshots of geometries at the phase transition for $N_3 = 6\mathrm{k}$. 
These snapshots hint at geometric features distinct from those seen in both phases: no tree-like features are present in the dual graph compared to the BP phase, while the geometries appear less like planar disks in appearance compared to those deep in the TT phase.

\section{Discussion}
Ordinary three-dimensional DT features two numerically well-understood phases, the crumpled phase (CR) in which triangulations typically have a macroscopic fraction of tetrahedra connected to a single vertex of huge degree, and the branched polymer (BP) phase where the geometry is tree-like. Adding spanning trees to the model effectively introduces Boltzmann weights for the geometries, modifying the partition function of the model.
However, the entropy of the spanning trees alone is seen not to be enough to alter the original phase diagram qualitatively. 
The introduction of interactions between the trees with the new coupling constants $\beta_{\mathrm{L}}$ and $\beta_{\mathrm{C}}$ radically changes the picture.
For increasingly negative $\beta_{\mathrm{L}}$ the tree-like features start dominating in the middle graph, which drives the crumpled phase to go through a phase transition. 
The new triple-tree (TT) phase occupies the corner of the phase diagram where all three couplings $\kappa_0$, $\beta_{\mathrm{L}}$ and $\beta_{\mathrm{C}}$ are negative. 
It borders the CR phase in the direction of increasing $\beta_{\mathrm{L}}$, the BP phase for increasing $\kappa_0$, while for increasing $\beta_{\mathrm{C}}$ yet another phase seems to appear, whose investigation we save for a forthcoming work. 
Regarding the nature of the transitions, we observed that BP-CR has the signature of a first-order transition, while the situation for TT-CR was not crystal clear: the phases show drastically different geometric features, but no scaling was observed. This could mean that we have not examined the appropriate order parameters, but a crossover phenomenon involving a separation of scales cannot be ruled out either. 
For the TT-BP transition our simulations indicate that it might well be continuous in nature. 
Histograms at the pseudocritical value of the coupling show no indication of first-order nature, while we observe accurate finite-size scaling with critical exponents that one would rather expect for a higher-order transition.

\subsection{Outlook}\label{sec:outlook}
In this work, we focused on the lower half of the parameter space, corresponding to $\beta_{\mathrm{C}} \leq 0$. 
This choice was motivated by preliminary explorations of the full parameter space, which indicated that the lower half exhibits comparatively more regular behaviour compared to the more complex geometric structures emerging in the upper half.
Determining whether these novel structures in the $\beta_{\mathrm{C}} > 0$ region represent distinct phases requires a more extensive and detailed analysis, which we defer to future work. 
Here we limit ourselves to highlighting some of the intriguing new phenomena observed.

\begin{figure}[h]
    \centering
    \includegraphics[width = 0.9\textwidth]{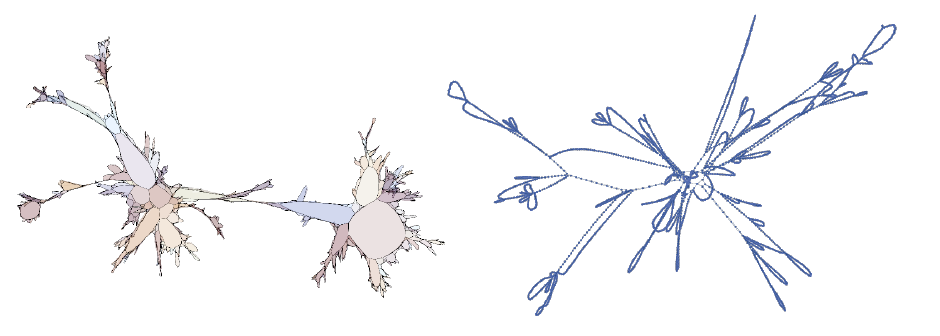}
    \caption{The dual graphs of typical random geometries at $\beta_{\mathrm{C}} = 3$ (left) and $\beta_{\mathrm{C}} = 10$ (right) along the edge $\kappa_0 = -10$ and $\beta_{\mathrm{L}} = -10$ of parameter space for $N_3 = 6\mathrm{k}$.}
    \label{fig:m10_m10_}
\end{figure}

For example, when we increase the value of $\beta_C$ while staying on the edge $\kappa_0 = -10$, $\beta_L = -10$ of the parameter space, some order parameters display two jumps as function of $\beta_C$. 
The dual graphs of typical random geometries at $\beta_{\mathrm{C}} = 3$ and $\beta_{\mathrm{C}} = 10$ are shown in Fig.~\ref{fig:m10_m10_}, hinting at the appearance of new structures.
In particular, we observe edges of increasingly large coordination number, corresponding in the images to long loops.

\begin{figure}[h]
    \centering
    \includegraphics[width = 0.45\textwidth]{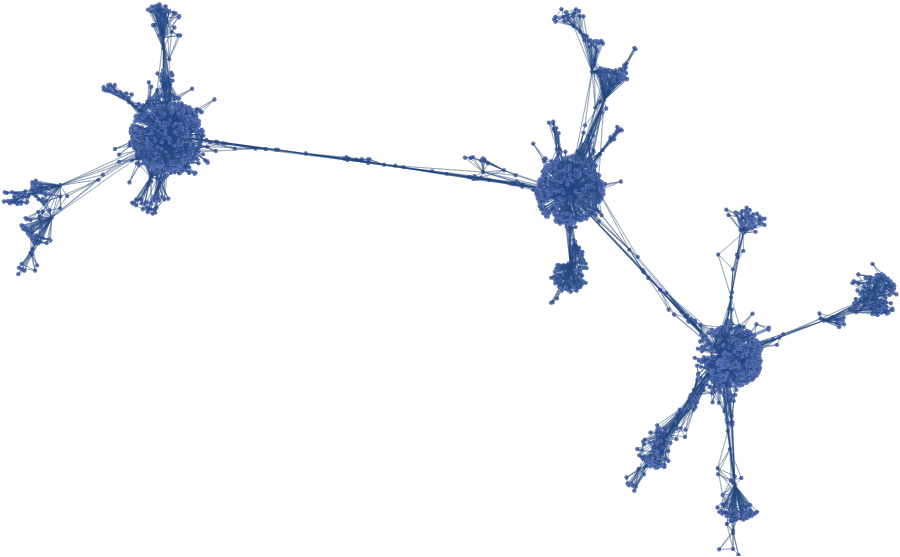}
    \includegraphics[width = 0.45\textwidth]{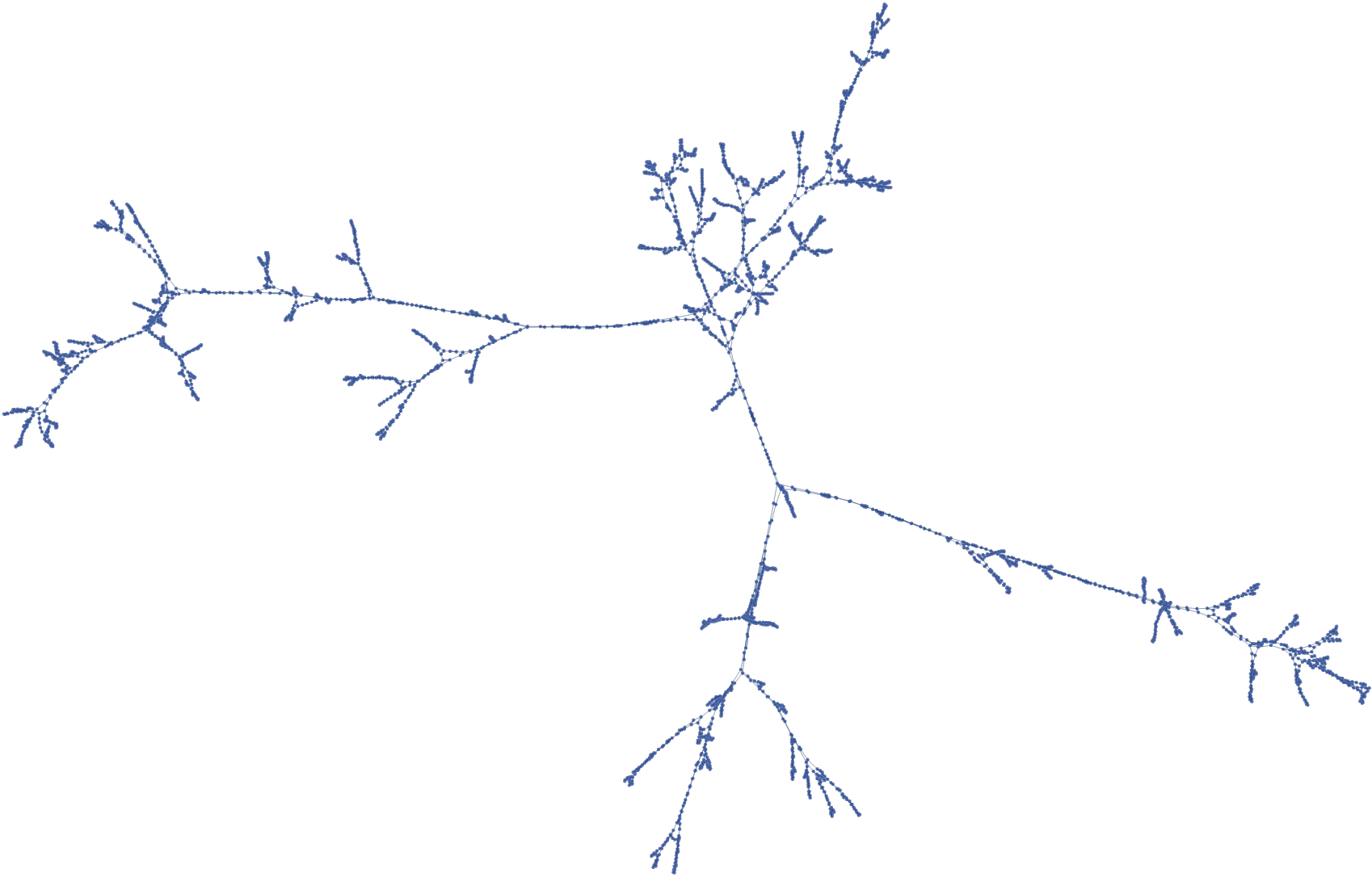}
    \includegraphics[width = 0.25\textwidth]{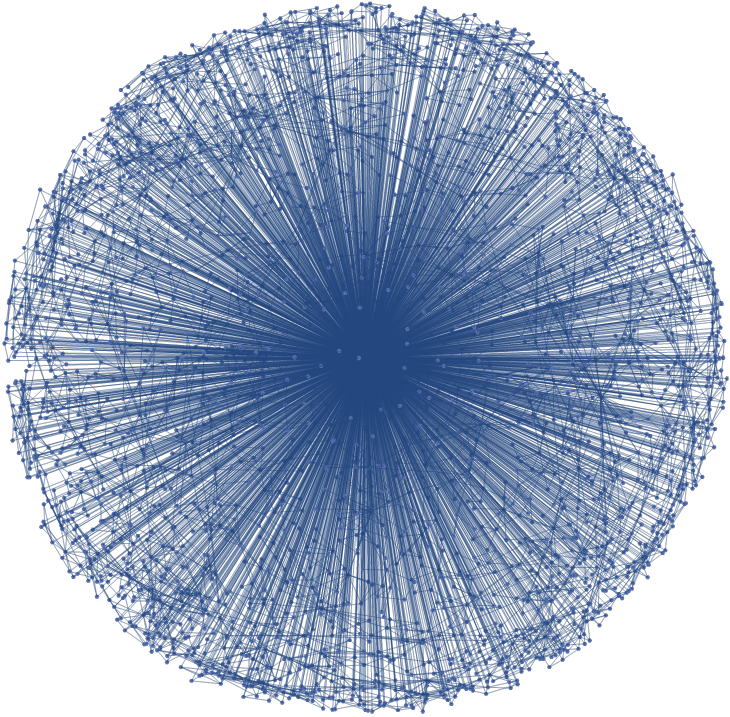}
    \includegraphics[width = 0.45\textwidth]{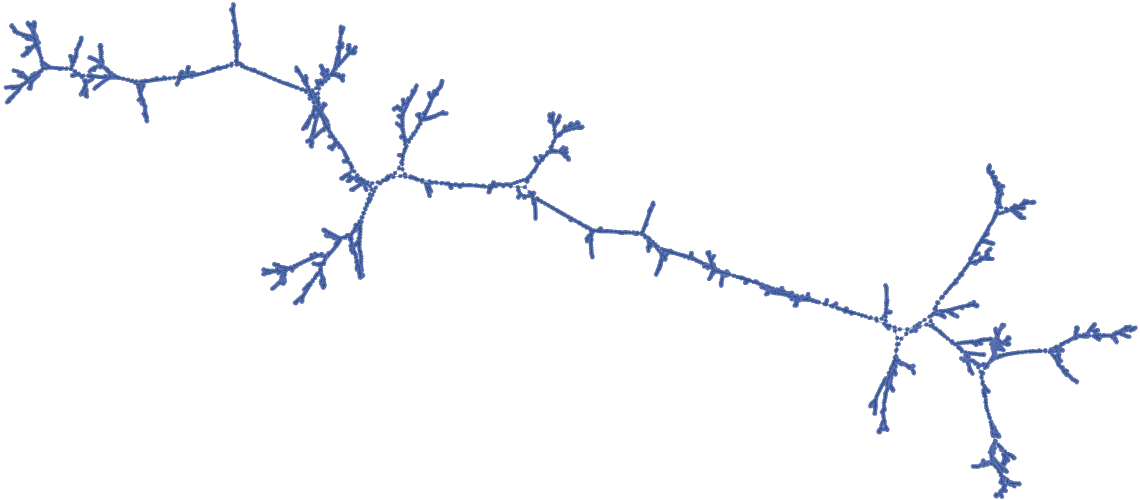}
    \caption{The vertex graph (left) and dual graph (right) of typical random geometries sampled at $\beta_{\mathrm{C}} = 3$ (top) and $\beta_{\mathrm{C}} = 10$ (bottom) along the edge $\kappa_0 = 10$ and $\beta_{\mathrm{L}}=-10$ of parameter space for $N_3 = 6\mathrm{k}$.}
    \label{fig:10_m10_3}
\end{figure}

At the other edge of the parameter space for $\kappa_0 = 10$ and $\beta_{\mathrm{L}}=-10$ and increasing $\beta_{\mathrm{C}}$ we find huge vertex coordination numbers. Here we see huge vertex coordination numbers, which reveal themselves as large balls on the vertex graphs shown in Fig.~\ref{fig:10_m10_3}.

\subsection*{Acknowledgements}
We thank Luca Lionni and Tom Gerstel for many discussions regarding particularly the triple-tree regime of this study.
This work is supported by the VIDI programme with project number VI.Vidi.193.048, which is financed by the Dutch Research Council (NWO).

\appendix
\section{Bounds}

\subsection{The bound $2 N_0 \leq N_3 + 6$}\label{sec:melonicbound}

Here we demonstrate that the melonic triangulations, which satisfy $2 N_0 = N_3 + 6$, are the restricted degenerate triangulations with the maximal number of vertices. 
To this end, let $M \geq 2$ and assume that $\mathcal{T}$ is a triangulation with $N_3 \leq M$ tetrahedra that maximizes $A := 2N_0 - N_3$.
We already know that $A\geq 6$, so we need to show that $A = 6$ and that $\mathcal{T}$ is necessarily a melonic triangulation.

To this end we observe that the average coordination number of all edges, i.e. the average number of triangles adjacent to an edge, in $\mathcal{T}$ is equal to 
\begin{equation*}
    \frac{3 N_2}{N_1} = \frac{6 N_3}{N_0+N_3} = \frac{4N_3}{N_3+ \frac{1}{3}A} < 4.
\end{equation*}
Hence $\mathcal{T}$ contains an edge $e$ with coordination number smaller than 4.
Coordination number 1 is not possible in a restricted degenerate triangulation.
If the coordination number of $e$ is 3, the triple of tetrahedra around $e$ contain 5 distinct vertices, and one can apply the inverse bistellar flip move to this triple.
The result is a restricted degenerate triangulation with the same number of vertices, but one fewer tetrahedron.
This is in contradiction with $A$ being maximal for $\mathcal{T}$.
Hence $e$, must have coordination number 2.
The pair of tetrahedra around $e$ are thus glued to each other along at least two of their triangles and share the same set of 4 vertices.
If they share exactly two triangles, then the inverse quadrangular pillow move can be applied to this pair.
The result is again a restricted degenerate triangulation with the same number of vertices but fewer tetrahedra, which is ruled out for the same reason as before.
If the pair of tetrahedra shares 3 triangles, then one can apply an inverse triangular pillow move to $\mathcal{T}$ to obtain a restricted degenerate triangulation with one fewer vertex and two fewer tetrahedra, which thus maximizes $2N_0 - N_3$ as well.
The only remaining possibility is that they share all 4 triangles, so that $\mathcal{T}$ is the unique triangulation with $N_3=2$ tetrahedra and $N_0=4$ vertices, for which $A = 6$.
Iterating this argument, we conclude that any maximizing triangulation $\mathcal{T}$ can be reduced to the last situation by repeated inverse triangular pillow moves.
Hence $A=6$ and $\mathcal{T}$ is precisely of the melonic type.

\subsection{The bound $N_0 + N_{\mathrm{L}} \leq N_3 + 3$}\label{sec:NLbound}

Let $(\mathcal{T},\mathcal{S},\mathcal{S}^*)$ be a tree-decorated triangulation.
To prove the bound on the number of loops $N_{\mathrm{L}}$ of the middle graph $\mathcal{G}(\mathcal{T},\mathcal{S},\mathcal{S}^*)$, it is convenient to examine the associated Apollonian triangulation $\mathcal{A}$, as illustrated in Fig.~\ref{fig:tree-tetrahedra}c.
It corresponds precisely to the boundary of the tetrahedron tree $\mathcal{S}^*$ when viewed as a gluing of $N_3$ tetrahedra of $\mathcal{T}$ according to the adjacency relations of $\mathcal{S}^*$.
The triangles of $\mathcal{T}$ that belong to the middle graph correspond to pairs of triangles in $\mathcal{A}$.
Moreover, the edges of $\mathcal{A}$ are partitioned into subsets such that each subset corresponds to an edge of $\mathcal{T}$.
The edges corresponding to the vertex tree $\mathcal{S}$ thus determine a subgraph $G \subset \mathcal{A}$ containing all $N_3 + 3$ vertices of $\mathcal{A}$ but only a subset of the edges, shown in green in the figure.
Let us denote by $E_G$ the number of edges of $G$, by $K_G$ the number of connected components of $G$ and by $F_G$ the number of faces of $G$, i.e.\ the number of connected components of the complement of $G$ in the plane.
Since $G$ has $N_3 + 3$ vertices, Euler's relation implies
\begin{equation}
    F_G = E_G + K_G - N_3 -2.
\end{equation}
Moreover, one may observe an edge of the middle graph corresponds precisely to two pairs consisting of a triangle of $\mathcal{A}$ and a neighboring edge of $\mathcal{A}$ that is not in $G$.
Since $\mathcal{A}$ has $3N_3 + 3$ edges in total, we find that the number of edges $N_{\mathrm{e}}$ of the middle graph is related to $E_G$ via
\begin{equation}
    N_{\mathrm{e}} = 3 N_3+3 - E_G.
\end{equation}
Combining with \eqref{eq:middlegraphrelation}, then yields
\begin{equation}\label{eq:NLGrelation}
    N_{\mathrm{L}} = N_{\mathrm{C}} - F_G + K_G - 1.
\end{equation}

Now we observe two inequalities. 
First, triangles in the same face of $G$ correspond to triangles in the middle graph that are connected.
This implies that $N_C \leq F_G$.
Second, upon appropriate identifications among the $N_3+3$ vertices of $G$ its adjacency must descend to the vertex tree $\mathcal{S}$.
Since the latter is connected, at least $K_G - 1$ vertex identifications must happen, so at most $N_3+3 - (K_G - 1)$ vertices remain in $\mathcal{S}$.
But we know $\mathcal{S}$ has $N_0$ vertices, so we have the two inequalities
\begin{equation}
    N_C \leq F_G, \qquad N_0 + K_G \leq N_3 + 4.
\end{equation}
Combining these with \eqref{eq:NLGrelation}, gives the claimed inequality 
\begin{equation}
    N_0 + N_{\mathrm{L}} \leq N_3 +3.
\end{equation}

\bibliographystyle{abbrvnat}
\bibliography{bib}

\begin{thebibliography}{42}
\providecommand{\natexlab}[1]{#1}
\providecommand{\url}[1]{\texttt{#1}}
\expandafter\ifx\csname urlstyle\endcsname\relax
  \providecommand{\doi}[1]{doi: #1}\else
  \providecommand{\doi}{doi: \begingroup \urlstyle{rm}\Url}\fi

\bibitem[Addario-Berry and Albenque(2017)]{addario-berry2017}
L.~Addario-Berry and M.~Albenque.
\newblock The scaling limit of random simple triangulations and random simple quadrangulations.
\newblock \emph{Ann. Probab.}, 45\penalty0 (5):\penalty0 2767--2825, 2017.
\newblock ISSN 0091-1798.
\newblock \doi{10.1214/16-AOP1124}.

\bibitem[Agishtein and Migdal(1991)]{agishtein1991three}
M.~Agishtein and A.~A. Migdal.
\newblock Three-dimensional quantum gravity as dynamical triangulation.
\newblock \emph{Modern Physics Letters A}, 6\penalty0 (20):\penalty0 1863--1884, 1991.

\bibitem[Ambj{\o}rn and Varsted(1992)]{Ambjoern_Three_1992}
J.~Ambj{\o}rn and S.~Varsted.
\newblock Three-dimensional simplicial quantum gravity.
\newblock \emph{Nuclear Physics B}, 373\penalty0 (2):\penalty0 557--577, 1992.

\bibitem[Ambj{\o}rn et~al.(1992)Ambj{\o}rn, Boulatov, Krzywicki, and Varsted]{Ambjoern_vacuum_1992}
J.~Ambj{\o}rn, D.~Boulatov, A.~Krzywicki, and S.~Varsted.
\newblock The vacuum in three-dimensional simplicial quantum gravity.
\newblock \emph{Physics Letters B}, 276\penalty0 (4):\penalty0 432--436, 1992.

\bibitem[Ambj{\o}rn et~al.(1993)Ambj{\o}rn, Burda, Jurkiewicz, and Kristjansen]{ambjorn1993three}
J.~Ambj{\o}rn, Z.~Burda, J.~Jurkiewicz, and C.~Kristjansen.
\newblock Three-dimensional simplicial quantum gravity coupled to ising matter.
\newblock \emph{Nuclear Physics B-Proceedings Supplements}, 30:\penalty0 771--774, 1993.

\bibitem[Ambjorn et~al.(1993)Ambjorn, Jurkiewicz, and Kristjansen]{Ambjorn:1992aw}
J.~Ambjorn, J.~Jurkiewicz, and C.~F. Kristjansen.
\newblock {Quantum gravity, dynamical triangulations and higher derivative regularization}.
\newblock \emph{Nucl. Phys. B}, 393:\penalty0 601--632, 1993.
\newblock \doi{10.1016/0550-3213(93)90075-Z}.

\bibitem[Ambjorn et~al.(1997)Ambjorn, Anagnostopoulos, Ichihara, Jensen, Kawamoto, Watabiki, and Yotsuji]{Ambjorn:1996kb}
J.~Ambjorn, K.~N. Anagnostopoulos, T.~Ichihara, L.~Jensen, N.~Kawamoto, Y.~Watabiki, and K.~Yotsuji.
\newblock {Quantum geometry of topological gravity}.
\newblock \emph{Phys. Lett. B}, 397:\penalty0 177--184, 1997.
\newblock \doi{10.1016/S0370-2693(97)00183-4}.

\bibitem[Ambj{\o}rn et~al.(2000)Ambj{\o}rn, Boulatov, Kawamoto, and Watabiki]{ambjorn2000recursive}
J.~Ambj{\o}rn, D.~Boulatov, N.~Kawamoto, and Y.~Watabiki.
\newblock Recursive sampling simulations of 3d gravity coupled to scalar fermions.
\newblock \emph{Physics Letters B}, 480\penalty0 (3-4):\penalty0 319--330, 2000.

\bibitem[Ambj\o{}rn et~al.(2001)Ambj\o{}rn, Jurkiewicz, and Loll]{ambjorn2001nonperturbative}
J.~Ambj\o{}rn, J.~Jurkiewicz, and R.~Loll.
\newblock Nonperturbative 3d lorentzian quantum gravity.
\newblock \emph{Phys. Rev. D}, 64:\penalty0 044011, Jul 2001.
\newblock \doi{10.1103/PhysRevD.64.044011}.

\bibitem[Ambjorn et~al.(2012{\natexlab{a}})Ambjorn, Barkley, and Budd]{Ambjorn:2011rs}
J.~Ambjorn, J.~Barkley, and T.~G. Budd.
\newblock {Roaming moduli space using dynamical triangulations}.
\newblock \emph{Nucl. Phys. B}, 858:\penalty0 267--292, 2012{\natexlab{a}}.
\newblock \doi{10.1016/j.nuclphysb.2012.01.010}.

\bibitem[Ambjorn et~al.(2012{\natexlab{b}})Ambjorn, Goerlich, Jurkiewicz, and Loll]{Ambjorn:2012jv}
J.~Ambjorn, A.~Goerlich, J.~Jurkiewicz, and R.~Loll.
\newblock {Nonperturbative Quantum Gravity}.
\newblock \emph{Phys. Rept.}, 519:\penalty0 127--210, 2012{\natexlab{b}}.
\newblock \doi{10.1016/j.physrep.2012.03.007}.

\bibitem[Ambjørn et~al.(1992)Ambjørn, Burda, Jurkiewicz, and Kristjansen]{AMBJORN1992253}
J.~Ambjørn, Z.~Burda, J.~Jurkiewicz, and C.~Kristjansen.
\newblock 3d quantum gravity coupled to matter.
\newblock \emph{Physics Letters B}, 297\penalty0 (3):\penalty0 253--260, 1992.
\newblock ISSN 0370-2693.
\newblock \doi{https://doi.org/10.1016/0370-2693(92)91258-B}.

\bibitem[Bernardi(2007)]{bernardi2007}
O.~Bernardi.
\newblock Bijective counting of tree-rooted maps and shuffles of parenthesis systems.
\newblock \emph{Electron. J. Combin.}, 14\penalty0 (1):\penalty0 Research Paper 9, 36, 2007.
\newblock \doi{10.37236/928}.

\bibitem[Bilke et~al.(1998{\natexlab{a}})Bilke, Burda, Krzywicki, Petersson, Tabaczek, and Thorleifsson]{bilke19984d}
S.~Bilke, Z.~Burda, A.~Krzywicki, B.~Petersson, J.~Tabaczek, and G.~Thorleifsson.
\newblock 4d simplicial quantum gravity: matter fields and the corresponding effective action.
\newblock \emph{Physics Letters B}, 432\penalty0 (3-4):\penalty0 279--286, 1998{\natexlab{a}}.

\bibitem[Bilke et~al.(1998{\natexlab{b}})Bilke, Burda, Krzywicki, Petersson, Tabaczek, and Thorleifsson]{bilke19984dgauge}
S.~Bilke, Z.~Burda, A.~Krzywicki, B.~Petersson, J.~Tabaczek, and G.~Thorleifsson.
\newblock 4d simplicial quantum gravity interacting with gauge matter fields.
\newblock \emph{Physics Letters B}, 418\penalty0 (3-4):\penalty0 266--272, 1998{\natexlab{b}}.

\bibitem[Bonzom et~al.(2011)Bonzom, Gurau, Riello, and Rivasseau]{bonzom2011critical}
V.~Bonzom, R.~Gurau, A.~Riello, and V.~Rivasseau.
\newblock Critical behavior of colored tensor models in the large n limit.
\newblock \emph{Nuclear Physics B}, 853\penalty0 (1):\penalty0 174--195, 2011.

\bibitem[Boulatov and Krzywicki(1991)]{Boulatov_phase_1991}
D.~Boulatov and A.~Krzywicki.
\newblock On the phase diagram of three-dimensional simplicial quantum gravity.
\newblock \emph{Modern Physics Letters A}, 6\penalty0 (32):\penalty0 3005--3014, 1991.

\bibitem[Boulatov et~al.(1986)Boulatov, Kazakov, Kostov, and Migdal]{BOULATOV1986641}
D.~Boulatov, V.~Kazakov, I.~Kostov, and A.~Migdal.
\newblock Analytical and numerical study of a model of dynamically triangulated random surfaces.
\newblock \emph{Nuclear Physics B}, 275\penalty0 (4):\penalty0 641--686, 1986.
\newblock ISSN 0550-3213.
\newblock \doi{https://doi.org/10.1016/0550-3213(86)90578-X}.

\bibitem[Broder(1989)]{broder1989generating}
A.~Z. Broder.
\newblock Generating random spanning trees.
\newblock In \emph{FOCS}, volume~89, pages 442--447, 1989.

\bibitem[Bruegmann and Marinari(1993)]{bruegmann19934d}
B.~Bruegmann and E.~Marinari.
\newblock 4d simplicial quantum gravity with a nontrivial measure.
\newblock \emph{Physical review letters}, 70\penalty0 (13):\penalty0 1908, 1993.

\bibitem[Brunekreef and N\'emeth(2022)]{Brunekreef:2022pns}
J.~Brunekreef and D.~N\'emeth.
\newblock {The phase structure and effective action of 3D CDT at higher spatial genus}.
\newblock \emph{JHEP}, 09:\penalty0 212, 2022.
\newblock \doi{10.1007/JHEP09(2022)212}.

\bibitem[Budd and Castro(2023)]{buddcastro}
T.~Budd and A.~Castro.
\newblock Scale-invariant random geometry from mating of trees: A numerical study.
\newblock \emph{Phys. Rev. D}, 107:\penalty0 026010, Jan 2023.
\newblock \doi{10.1103/PhysRevD.107.026010}.

\bibitem[Budd and Lionni(2022)]{budd2022family}
T.~Budd and L.~Lionni.
\newblock A family of triangulated 3-spheres constructed from trees.
\newblock 2022.
\newblock arXiv preprint arXiv:2203.16105.

\bibitem[Catterall et~al.(1995)Catterall, Kogut, and Renken]{Catterall_Entropy_1995}
S.~Catterall, J.~Kogut, and R.~Renken.
\newblock Entropy and the approach to the thermodynamic limit in three-dimensional simplicial gravity.
\newblock \emph{Physics Letters B}, 342\penalty0 (1-4):\penalty0 53--57, 1995.

\bibitem[Coumbe and Laiho(2015)]{Coumbe:2014nea}
D.~Coumbe and J.~Laiho.
\newblock {Exploring Euclidean Dynamical Triangulations with a Non-trivial Measure Term}.
\newblock \emph{JHEP}, 04:\penalty0 028, 2015.
\newblock \doi{10.1007/JHEP04(2015)028}.

\bibitem[Duplantier et~al.(2021)Duplantier, Miller, and Sheffield]{duplantier2021}
B.~Duplantier, J.~Miller, and S.~Sheffield.
\newblock Liouville quantum gravity as a mating of trees.
\newblock \emph{Ast\'{e}risque}, \penalty0 (427):\penalty0 viii+257, 2021.
\newblock ISSN 0303-1179.
\newblock \doi{10.24033/ast}.

\bibitem[Eisenstat()]{dtree}
D.~Eisenstat.
\newblock dtree: dynamic trees à la carte (c++).
\newblock \url{https://www.davideisenstat.com/dtree/}.

\bibitem[Ferri et~al.(1986)Ferri, Gagliardi, and Grasselli]{ferri1986graph}
M.~Ferri, C.~Gagliardi, and L.~Grasselli.
\newblock A graph-theoretical representation of pl-manifolds—a survey on crystallizations.
\newblock \emph{Aequationes mathematicae}, 31:\penalty0 121--141, 1986.

\bibitem[Gerstel(2023)]{gerstel}
T.~Gerstel.
\newblock Simulating quantum gravity using triangles and trees.
\newblock Master's thesis, Radboud University, 2023.
\newblock URL \url{https://surfdrive.surf.nl/files/index.php/s/dpHByvWQOD8B9GJ/download?path=%2FGravity&files=2023%20-%20Simulating%20Quantum%20Gravity%20-%20Tom%20Gerstel.pdf}.

\bibitem[Hagura et~al.(1998)Hagura, Tsuda, and Yukawa]{Hagura_Phases_1998}
H.~Hagura, N.~Tsuda, and T.~Yukawa.
\newblock Phases and fractal structures of three-dimensional simplicial gravity.
\newblock \emph{Physics Letters B}, 418\penalty0 (3-4):\penalty0 273--283, 1998.

\bibitem[Holm et~al.(2001)Holm, de~Lichtenberg, and Thorup]{polylog}
J.~Holm, K.~de~Lichtenberg, and M.~Thorup.
\newblock Poly-logarithmic deterministic fully-dynamic algorithms for connectivity, minimum spanning tree, 2-edge, and biconnectivity.
\newblock \emph{J. ACM}, 48:\penalty0 723--760, 07 2001.
\newblock \doi{10.1145/276698.276715}.

\bibitem[Hotta et~al.(1998)Hotta, Izubuchi, and Nishimura]{Hotta_Multicanonical_1998}
T.~Hotta, T.~Izubuchi, and J.~Nishimura.
\newblock Multicanonical simulation of 3d dynamical triangulation model and a new phase structure.
\newblock \emph{Nuclear Physics B}, 531\penalty0 (1-3):\penalty0 446--458, 1998.

\bibitem[Kawamoto et~al.(1992)Kawamoto, Kazakov, Saeki, and Watabiki]{Kawamoto:1991xd}
N.~Kawamoto, V.~A. Kazakov, Y.~Saeki, and Y.~Watabiki.
\newblock {Fractal structure of two-dimensional gravity coupled to D = -2 matter}.
\newblock \emph{Phys. Rev. Lett.}, 68:\penalty0 2113--2116, 1992.
\newblock \doi{10.1103/PhysRevLett.68.2113}.

\bibitem[Kazakov et~al.(1985)Kazakov, Kostov, and Migdal]{kazakov1985critical}
V.~A. Kazakov, I.~Kostov, and A.~Migdal.
\newblock Critical properties of randomly triangulated planar random surfaces.
\newblock \emph{Physics Letters B}, 157\penalty0 (4):\penalty0 295--300, 1985.

\bibitem[Le~Gall(2013)]{legall2013}
J.-F. Le~Gall.
\newblock Uniqueness and universality of the {B}rownian map.
\newblock \emph{Ann. Probab.}, 41\penalty0 (4):\penalty0 2880--2960, 2013.
\newblock ISSN 0091-1798.
\newblock \doi{10.1214/12-AOP792}.

\bibitem[Mullin(1967)]{mullin1967}
R.~C. Mullin.
\newblock On the enumeration of tree-rooted maps.
\newblock \emph{Canadian J. Math.}, 19:\penalty0 174--183, 1967.
\newblock ISSN 0008-414X.
\newblock \doi{10.4153/CJM-1967-010-x}.

\bibitem[Pemantle(2004)]{pemantle2004uniform}
R.~Pemantle.
\newblock Uniform random spanning trees.
\newblock 2004.

\bibitem[Sleator and Tarjan(1981)]{sleator1981data}
D.~D. Sleator and R.~E. Tarjan.
\newblock A data structure for dynamic trees.
\newblock In \emph{Proceedings of the thirteenth annual ACM symposium on Theory of computing}, pages 114--122, 1981.

\bibitem[Thorleifsson(1999)]{Thorleifsson:1998nb}
G.~Thorleifsson.
\newblock {Three-dimensional simplicial gravity and degenerate triangulations}.
\newblock \emph{Nucl. Phys. B}, 538:\penalty0 278--294, 1999.
\newblock \doi{10.1016/S0550-3213(98)00679-8}.

\bibitem[Tseng()]{Tom-Tseng}
T.~Tseng.
\newblock {Dynamic Connectivity}.
\newblock \url{https://github.com/tomtseng/dynamic-connectivity-hdt}.
\newblock GitHub repository.

\bibitem[Weinberg(1980)]{Weinberg:1980gg}
S.~Weinberg.
\newblock \emph{Ultraviolet divergences in quantum theories of gravitation}, pages 790--831.
\newblock 1980.

\bibitem[Wilson(1974)]{Wilson:1974sk}
K.~G. Wilson.
\newblock {Confinement of Quarks}.
\newblock \emph{Phys. Rev. D}, 10:\penalty0 2445--2459, 1974.
\newblock \doi{10.1103/PhysRevD.10.2445}.

\end{thebibliography}

\end{document}